\documentclass[fleqn,usenatbib]{mnras}

\usepackage{newtxtext,newtxmath}

\usepackage[T1]{fontenc}

\DeclareRobustCommand{\VAN}[3]{#2}
\let\VANthebibliography\thebibliography
\def\thebibliography{\DeclareRobustCommand{\VAN}[3]{##3}\VANthebibliography}

\usepackage{graphicx}	
\usepackage{amsmath}	
\usepackage{siunitx}
\usepackage{nicefrac}
\usepackage{float}
\usepackage{orcidlink}
\usepackage[normalem]{ulem} 

\newcommand{\mcg}{MCG\,$+$08$-$11$-$11}	            
\newcommand{\swift}{\textit{Swift}}	                
\newcommand{\nustar}{\textit{NuSTAR}}	            
\newcommand{\javelin}{\textsc{javelin}}	            
\newcommand{\kynreverb}{\textsc{KYNreverb}}	        
\newcommand{\psresp}{\textsc{psresp}}	            
\newcommand{\rg}{$R_\mathrm{g}$}
\newcommand{\msun}{M$_{\odot}$}
\newcommand{\me}{$\dot{m}_\mathrm{E}$}
\newcommand{\xmm}{\textit{XMM-Newton}}

\makeatletter
\newcommand{\Mc}{}
\DeclareRobustCommand{\Mc}{%
  M%
  \raisebox{\dimexpr\fontcharht\font`M-\height}{%
    \check@mathfonts\fontsize{\sf@size}{0}\selectfont
    {c}%
  }%
}
\makeatother

\graphicspath{{Figures/}}

\title[Intensive continuum reverberation mapping of MCG+08-11-11]{Intensive X-Ray/UVOIR continuum reverberation mapping of the Seyfert AGN \mcg}

\author[D. Kynoch et al.]{
D.\ Kynoch,$^{\orcidlink{0000-0001-8638-3687}~{1,2}}$\thanks{E-mail: daniel.kynoch@port.ac.uk}
I.~M.\ {\Mc}Hardy,$^{\orcidlink{0000-0002-0151-2732}~{1}}$
E.~M.\ Cackett,$^{\orcidlink{0000-0002-8294-9281}~{3}}$
J.\ Gelbord,$^{\orcidlink{0000-0001-9092-8619}~{4}}$
J.~V.\ Hern\'{a}ndez Santisteban,$^{\orcidlink{0000-0002-6733-5556}~{5}}$
\newauthor
K.\ Horne,$^{\orcidlink{0000-0003-1728-0304}~{5}}$
J.~A.\ Miller,$^{\orcidlink{0000-0001-8475-8027}~{6,3}}$
H.\ Netzer,$^{\orcidlink{0000-0002-6766-0260}~{7}}$
C.\ Done,$^{\orcidlink{0000-0002-1065-7239}~{8}}$
R.~Edelson~$^{\orcidlink{0000-0001-8598-1482}~{9}}$,
M.~M.\ Fausnaugh,$^{\orcidlink{0000-0002-9113-7162}~{10}}$
\newauthor
M.~R.\ Goad,$^{\orcidlink{0000-0002-2908-7360}~{11}}$
B.~M.\ Peterson,$^{\orcidlink{0000-0001-6481-5397}~{12}}$ 
F.~M.\ Vincentelli$^{\orcidlink{0000-0002-1481-1870}~{13,1}}$ \\
$^{1}$School of Physics and Astronomy, University of Southampton, University Road, Southampton, SO17 1BJ, UK\\
$^{2}$Institute of Cosmology and Gravitation, University of Portsmouth, Burnaby Road, Portsmouth, PO1 3FX, UK\\
$^{3}$Department of Physics and Astronomy, Wayne State University, 666 W. Hancock St, Detroit, MI, 48201, USA\\
$^{4}$Spectral Sciences Inc., 4 Fourth Avenue, Burlington, MA 01803, USA\\
$^{5}$SUPA School of Physics and Astronomy, University of St Andrews, North Haugh, St Andrews, KY16 9SS, Scotland, UK\\
$^{6}$Texas A\&M University, Department of Physics \& Astronomy, 400 Bizzell St, College Station, TX 77845, USA\\
$^{7}$School of Physics and Astronomy, Tel Aviv University, Tel Aviv 69978, Israel\\
$^{8}$Centre for Extragalactic Astronomy, Department of Physics, University of Durham, South Road, Durham, DH1 3LE, UK\\
$^{9}$ Eureka Scientific Inc., 2452 Delmer St. Suite 100, Oakland, CA 94602, USA\\
$^{10}$ Department of Physics \& Astronomy, Texas Tech University, Lubbock TX, 79409-1051, USA\\
$^{11}$Department of Physics and Astronomy, College of Science and Engineering, University of Leicester, University Road, Leicester, LE1 7RH, UK\\
$^{12}$Retired\\
$^{13}$INAF Istituto di Astrofisica e Planetologia Spaziali, Via del Fosso del Cavaliere 100, 00133 Roma, Italy\\
}

\date{Accepted XXX. Received YYY; in original form ZZZ}

\pubyear{2025}

\begin{document}
\label{firstpage}
\pagerange{\pageref{firstpage}--\pageref{lastpage}}
\maketitle

\begin{abstract}
We present results from intensive ($\times3$ daily), three-month-long X-ray, UV and optical monitoring of the bright Seyfert active galactic nucleus (AGN) \mcg\ with \swift, supported by optical–infrared ground-based monitoring.  The 12 resultant, well-sampled, lightcurves are highly correlated; in particular, the X-ray–to–UV correlation $r_\mathrm{max}=0.85$ is, as far as we know, the highest yet recorded in a Seyfert galaxy.  The lags increase with wavelength, as expected from reprocessing of central high-energy emission by surrounding material.  Our lag spectrum is much shallower than that obtained from an optical monitoring campaign conducted a year earlier when \mcg\ was $\approx4$ times brighter.  After filtering out long-term trends in the earlier optical lightcurves we recover shorter lags consistent with our own – demonstrating concurrent reverberation signals from different spatial scales and the luminosity dependence of the measured lags.  We use our lag spectrum to test several physical models, finding that disc reprocessing models cannot account for the observed `excess’ lags in the $u$ and $r$--$i$-bands that are highly indicative of the Balmer and Paschen continua produced by reprocessing in the broad line region (BLR) gas.  The structure seen in both the variable (rms) and lag spectra, and the large time delay between X-ray and UV variations ($\approx2$~days) all suggest that the BLR is the dominant reprocessor.  The hard X-ray spectrum ($\Gamma\approx1.7$) and faint, red, UV-optical spectrum both indicate that the Eddington accretion ratio is low: $\dot{m}_\mathrm{E}\sim0.03$.  The bolometric luminosity then requires that the black hole mass is substantially greater than current reverberation mapping derived estimates. 
\end{abstract}




\section{Introduction}
Determining the inner structure of Active Galactic Nuclei (AGN), and how that structure might vary with parameters such as black hole mass and accretion rate, is a topic of major importance in astrophysics. However with the exceptions of M87 and Sgr A*, whose very inner regions have been mapped by very high resolution radio interferometry (\citealt{EHT19,EHT22}), almost all other AGN are too small for their structures to be determined by direct imaging. A more widely applicable indirect method of determining AGN inner structures is `Reverberation Mapping' (RM: \citealt{Bahcall72}; \citealt{Blandford82}). Under the assumption that low energy photons are produced by reprocessing of high energy photons from near the black hole, by surrounding material, the structure of that material can be deduced from the time lags between wavebands.

RM was first envisaged as a means to probe the size scale and geometry of line-emitting gas (i.e.\ the broad line region or BLR gas) by measuring the delayed response of emission lines to variations in the photoionising UV continuum, assumed to come from the inner accretion disc, very close to the black hole \cite[e.g.][]{Peterson93,Peterson04}.
More recently, high-cadence photometric monitoring campaigns have concentrated on measuring the lags between the continuum X-ray and UV, optical and infrared (UVOIR) wavebands \cite[e.g.][and many others]{Shappee14,McHardy14,Edelson15,McHardy18,Cackett18,Edelson19,Cackett21}. 
The lags have been compared with the expectations from reprocessing on a variety of surrounding structures.
Until recently, the standard paradigm has been that the dominant reprocessor was 
the accretion disc around the black hole, usually assumed to have the temperature profile described by \cite{SS73} which depends on the black hole mass ($M_\mathrm{BH}$), spin ($a_\star$) and accretion rate (which is commonly expressed as the ratio between an AGN's bolometric luminosity and its Eddington luminosity, i.e.\ $\dot{m}_\mathrm{E}\equiv L_\mathrm{bol}/L_\mathrm{Edd}$).
However, although the lags ($\tau$) usually increase with wavelength ($\lambda$) roughly in agreement with the expectation of disc reprocessing of $\tau \propto \lambda^{4/3}$ \citep{Cackett07}, there are a number of discrepancies. 

Firstly, the lags should depend on the size and structure of the central X-ray source. However, under the assumption that this source can be approximated by a lamppost geometry, with a height of $\sim10$~\rg, similar to that deduced from X-ray intraband reverberation \cite[e.g.][]{Emmanoulopoulos14}, the observed lags usually exceed the theoretical expectations, leading to speculation that perhaps the disc is clumpy, not smooth \citep{McHardy14}. Numerical disc modelling \cite[e.g.][]{McHardy14, McHardy18} usually leads to smaller discrepancies (factor $\sim2$) than analytic formulae \cite[factor 3 or greater, e.g.][]{Fausnaugh16, Edelson19}. 
More detailed disc modelling, including General Relativistic effects, larger X-ray source heights ($>10$~\rg), and a large disc colour temperature correction, suggest that there is no discrepancy (e.g.\ \citealt{Kammoun21}) - although these models have not yet been fully explored. 

Secondly, the lag between the X-rays and the near UV is usually factors of at least 5 greater than the lag expected from extrapolation of the lag spectrum between the optical and UV down to X-ray wavelengths \cite[e.g.][]{McHardy18}. However, that extrapolation is based on the assumption that a relationship of the form $\tau \propto \lambda^{4/3}$ describes the true shape of the lag spectrum. It has been shown that if there is an edge to the reprocessing region of the disc, perhaps caused by a disc wind arising at large radii which obscures the outer part of the disc, then the lag spectrum will flatten at long wavelengths \citep{McHardy23}. Thus an extrapolation of the long wavelength lag spectrum to short wavelengths will naturally lead to an apparent X-ray to UV lag excess.  

Thirdly, although the UVOIR bands are always well-correlated with each other, the correlation with the X-ray band is much weaker and sometimes no UV-X-ray correlation is found (e.g.\ \citealt{Kara21}). For example, in a study of 21 AGN with \swift\ X-ray and UV data, a statistically significant correlation between the X-ray and UV lightcurves was found in only 9 AGN \citep{Buisson17}. However \cite{McHardy14} and \cite{Pahari20} noted that removal of long term trends in lightcurves, as recommended by \cite{Welsh99}, can lead to stronger correlations and a much clearer detection of a reprocessing signature in the lags.
Also \cite{Panagiotou22} showed that a corona with varying height will reduce the X-ray/UVOIR correlation strength whilst maintaining the basic principles of the lamppost model. \cite{GD17} suggested that an inflated inner disc might scatter and delay the \textcolor{black}{hard} X-ray photons, which are then reprocessed into \textcolor{black}{soft X-ray}/far-UV photons which then illuminate the rest of the disc. Thus an additional delay is added to the \textcolor{black}{hard} X-ray/UVOIR lag and the \textcolor{black}{hard} X-ray/UVOIR correlation is diluted.

Fourthly, the lags, particularly in the $u$-band \citep[e.g.][]{Cackett18} but also, to a lesser extent, in the $i$-band \citep[e.g.][]{Miller23}, are larger than expected compared to the lags in surrounding bands. These lags probably arise from a contribution to the light in those wavebands from Balmer and Paschen continua respectively, caused by reprocessing in the more distant BLR \citep{KG01, KG19}. 
Frequency-resolved lag analyis also suggests a response originating from larger size scales, consistent with the BLR (\citealt{Cackett22}; \citealt{Lewin23}). The questions now are: what fraction of the reprocessed light comes from the BLR (possibly all?: \citealt{Netzer22}); and, does that fraction vary between AGN?

There are also other questions concerning reprocessing from a disc. For example, why are the UV/optical lightcurves so smooth, compared to the X-rays? For MR\,2251-178 \cite{Arevalo08} show that an X-ray source height of 100~\rg\, is required to produce smooth enough optical lightcurves. However such a height leads to larger X-ray to optical lags than are observed.
The more rapidly varying X-ray lightcurves also lead to low peak X-ray/optical cross-correlation coefficients, although some smoothing of the X-ray lightcurves can increase the peak coefficient \citep{Cackett20}.

To be able to discriminate between different physical models for the origin of the variability, and hence obtain a good understanding of the inner structures of AGN via RM, it is necessary to measure the lags between wavebands with high accuracy. Measurements with large uncertainty are not really very useful. Thus it is necessary to observe AGN which are very bright across all wavebands from X-ray to near-IR and which have established patterns of variability showing large amplitude variations on timescales which are well suited to the existing observational facilities. The target of the present investigation, \mcg, fulfils these criteria extremely well, as we describe below.  

\subsection{The source \mcg}\label{sec:mcg}
The subject of this study is the bright, nearby ($z=0.0205$) Seyfert galaxy \mcg\ (also known as UGC\,3374).
It contains an X-ray bright nucleus and was the first Seyfert galaxy to be discovered by its X-ray emission (\citealt{Ward77}).
It is known to be strongly variable over all X-ray, UV and optical wavebands. 

In several spectroscopic observations made since 2004 \mcg\ has been shown to have a flat X-ray spectrum ($\Gamma\approx1.8$, estimated over various energy bands between 0.5 and 150~keV: \citealt{Matt06}; \citealt{Bianchi10}; \citealt{Patrick12}; \citealt{Murphy14} and \citealt{Tortosa18}).
\cite{Matt06} measured the X-ray flux (2--10~keV, unabsorbed) as $4.63\times10^{-11}$~erg\,s$^{-1}$\,cm$^{-2}$, corresponding to a luminosity $\log(L_\mathrm{X}/\mathrm{erg\,s^{-1}})=43.6$.
\mcg\ is bright in even hard X-rays (14--195~keV) and has been detected by \swift-BAT, with its luminosity reported as $\log(L_\mathrm{X}/\mathrm{erg\,s^{-1}})=44.13$ (\citealt{Lien25})\footnote{\swift\ BAT 157-month hard X-ray survey: \url{https://swift.gsfc.nasa.gov/results/bs157mon/}}.
It is known to be variable in X-rays over a range of timescales: significant variability of its X-ray flux over a period of 18 months was detected by the \textit{Ariel\,V} Sky Survey Instrument (\citealt{Cooke78}) and modest variability on shorter timescales ($\sim10$~per cent variability in 38~ks) was detected in an \textit{XMM-Newton} observation in 2004 (\citealt{Matt06}).

In the optical, \mcg\ is a bright ($\mathrm{V}\approx14.2$~mag) AGN, spectroscopically classified as a Seyfert 1.5 (having broad and narrow emission lines of comparable strength: \citealt{Osterbrock82}).
Its high optical brightness and variability have made the source an ideal target for both photometric and spectroscopic RM campaigns (e.g.\ \citealt{Sergeev05}; \citealt{Fausnaugh17,Fausnaugh18} and \citealt{Fian23}).
\cite{Fausnaugh17} reported a black hole mass of $2.82 \times 10^{7}$~\msun, derived from optical line reverberation, with an optical continuum-H\,$\upbeta$ lag of $\approx15$ days. They note a factor of 3 uncertainty in the mass.
The H\,$\upbeta$ lag is in excellent agreement with the expectation based on its 5100~\si\angstrom\ luminosity and the BLR radius-luminosity relation of \cite{Bentz13}.
\cite{Ma23} determine a very similar lag for H\,$\upalpha$ ($\approx17$~days) from photometric monitoring data.
The Eddington accretion ratio, \me, is less certain. From the 5100~\si\angstrom\ luminosity, and assuming a bolometric correction factor of 10 and a mass of $2.82 \times 10^{7}$~\msun, \cite{Fausnaugh18} report $\dot{m}_\mathrm{E}=0.054$. 
However, from {\sc cream} modelling of the optical lightcurves, the same authors report $\dot{m}_\mathrm{E}=0.234$. We return to this point in Sec.~\ref{sec:sed}.

\mcg\ has been the subject of previous optical continuum RM studies which show it to be highly variable on $\sim$few day to month timescales across all optical bands.
\cite{Sergeev05} presented lag measurements between the B and V, R and I bands based on a $\sim2$-year campaign beginning in 2001, with 69 sets of observations and typical sampling of $\sim3$~days. They find roughly wavelength-dependent lags with the R-band lagging the B-band by about 5~days.
\cite{Fausnaugh18} measured lags between the $ugriz$ and 5100~\si\angstrom\ bands. Their total number of observations is slightly fewer than those of \cite{Sergeev05}, but with better sampling of $\sim1$~day. They also find roughly wavelength-dependent lags, but shorter than those of \citeauthor{Sergeev05}, with the $i$ band lagging the $g$ band by $\sim1.5$~days.
Most recently, \cite{Fian23} presented results from optical photometric monitoring of \mcg\ with approximately nightly cadence over six months in 2019-20 with the C18 telescope at the Wise Observatory.
They robustly detected wavelength-dependent optical interband lags of up to seven days (at 8025~\si\angstrom\ relative to 4250~\si\angstrom).
They noted that the measured wavelength dependence of these lags, $\tau\propto\lambda^{4.74}$, is very much steeper than the $\tau\propto\lambda^{\nicefrac{4}{3}}$ predicted for a standard thin accretion disc.
Furthermore, the long lags imply an accretion disc $\sim3$--7 times larger than predicted by standard, thin disc theory.

The observations discussed above show that \mcg\ is indeed a very bright AGN across all optical wavebands and that with a more intense observing campaign, it was very likely that good enough lags could be measured to allow for accurate modelling.
However, none of those studies included contemporaneous X-ray observations, which are essential for proper modelling of the emission geometry, since in the disc reverberation paradigm it is the X-rays which drive short-timescale variability at longer wavelengths. 
In this paper, we present the results from just such a RM campaign, covering a very wide range of wavelengths from X-rays to near-IR, over a 3-month period, with much more intensive sampling (better than twice daily) than previous studies.

\textcolor{black}{This paper is structured as follows. 
In Section~\ref{sec:observations} we describe the observations of \mcg, including our new \swift, LCO and Zowada data.
In Section~\ref{sec:timing} we perform timing analyses to assess the target's variability properties and determine its lag spectrum, $\tau(\lambda)$.
In Section~\ref{sec:sed} we investigate the multiwavelength spectral energy distribution (SED) of the source and estimate its Eddington accretion ratio, $\dot{m}_\mathrm{E}$.
In Section~\ref{sec:models}, informed by our SED and $\dot{m}_\mathrm{E}$ estimate, we model $\tau(\lambda)$ with thin-disc (\ref{sec:kyn}), flared-disc (\ref{sec:bowl}) and radiation-pressure-confined cloud (\ref{sec:blr}) models.
Finally, in Section~\ref{sec:discussion} we discuss our findings, and in Section~\ref{sec:conclusions} we summarise our conclusions.}

\begin{figure*}
    \centering
    \includegraphics[width=0.9\textwidth]{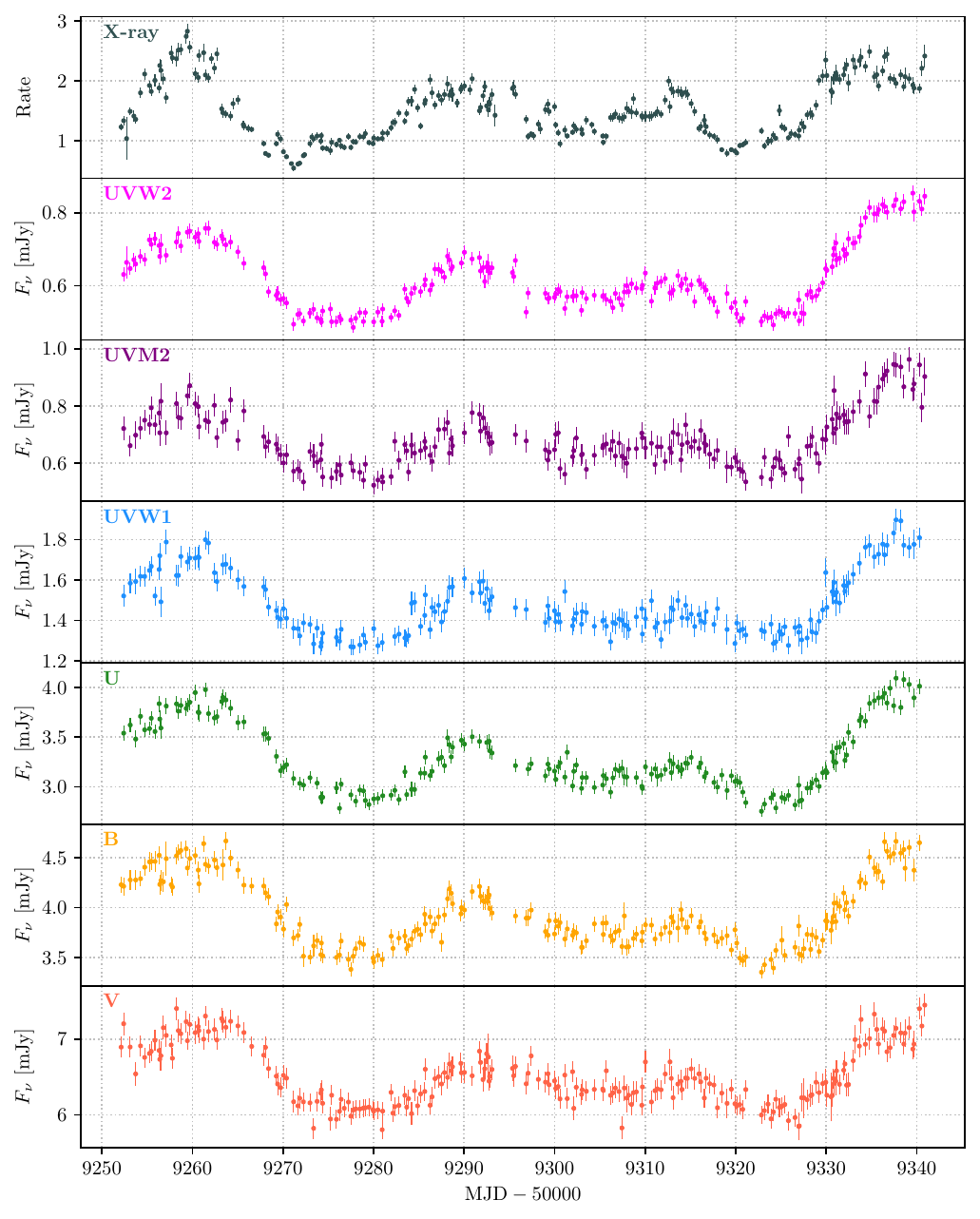}
    \caption{\swift\ XRT (top panel) and UVOT (lower panels) lightcurves from the three-month period between 2021 February 5 and May 7.}
    \label{fig:lcs}
\end{figure*}

\begin{figure*}
    \centering
    \includegraphics[width=0.9\textwidth]{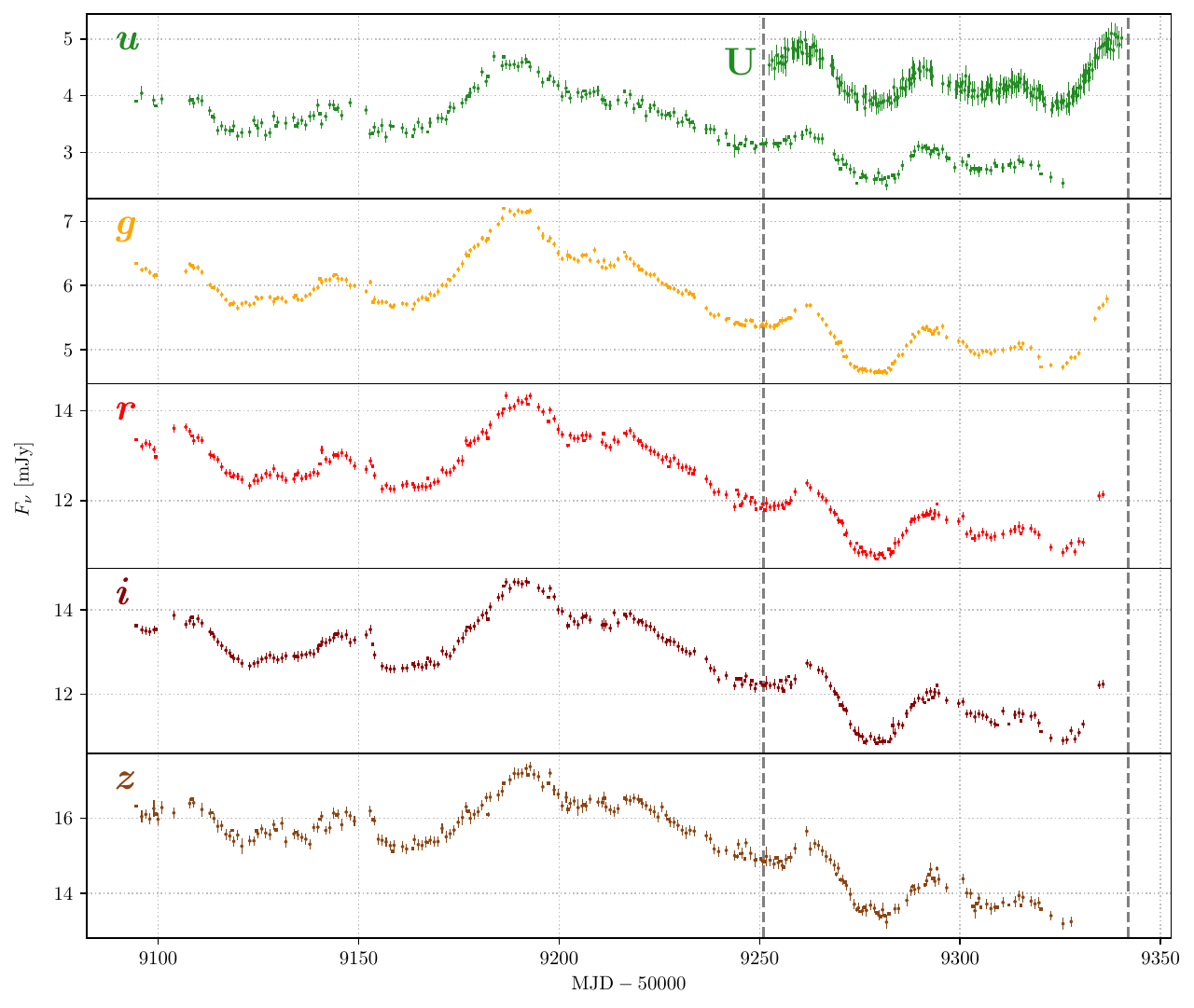}
    \caption{The combined Zowada and LCO lightcurves from the eight months between 2020 September and 2021 May.
    The beginning and end of the intensive three-month \swift\ monitoring period are indicated by vertical grey dashed lines.
    The \swift\ UVOT U-band lightcurve is shown in the top panel (offset by 1~mJy) for comparison.}
    \label{fig:optlcs}
\end{figure*}

\section{Observations}\label{sec:observations}
\subsection{2020-21 Continuum Reverberation Mapping Campaign}
\textcolor{black}{The structural conclusions regarding \mcg\ which we present later in this paper (Sections~\ref{sec:sed} and ~\ref{sec:models}) are based on timing analysis (Section~\ref{sec:timing}) of lightcurves obtained during} a major, coordinated monitoring campaign on \mcg\ conducted between 2020 January and 2021 November. \textcolor{black}{In this Section we describe these data.}

The principal observations were obtained by the Neil Gehrels Swift Observatory (hereafter \swift) in a period of intensive monitoring between 2021 February and May (Section~\ref{sec:data-swift}), although the source has been regularly monitored by \swift\ with low cadence for several months both before and after this period.
The \swift\ observations were supported by ground-based optical-IR photometric monitoring conducted by the Las Cumbres Observatory network (Section~\ref{sec:data-lco}) and the Dan Zowada Memorial Observatory (Section~\ref{sec:data-zowada}).
Regular, ground-based observations were conducted over the whole 2020-21 season when the target was at sufficiently low airmass during the night; the last three months of the ground observing seasons coincided with the intensive \swift\ monitoring period.
In this work we have also made use of archival data from the \nustar\ and \xmm\ Observatories, as described in Section~\ref{sec:data-nustar} and \ref{sec:data-xmm}, respectively.
A forthcoming paper (Keeley et al., in preparation) will present an analysis of the longer-term variability of \mcg.
In this paper we focus on the three-month period of intensive \swift\ monitoring.

\label{sec:data}
\subsubsection{Neil Gehrels Swift Observatory}
\label{sec:data-swift}
The \swift\ satellite is equipped with an X-ray Telescope (XRT: \citealt{Burrows05}) and Ultra-Violet/Optical Telescope (UVOT: \citealt{Roming05}; \citealt{Poole08}).  
As summarised in Table~\ref{tab:obs}, \mcg\ was intensively monitored by \swift\ between 2021 February 7 and May 7, with three visits per day during this three-month period (Cycle 17 proposal 1720061; PI {\Mc}Hardy).
During each visit, measurements were made in X-rays (0.3--10~keV) as well as six UV/optical bands: UVW2, UVM2, UVW1, U, B and V (see Fig~\ref{fig:lcs}).
The data were reduced using tools included in HEAsoft (\citealt{Nasa14}); the procedure is described in \cite{Edelson15}.

In addition to the X-ray lightcurve, we obtained a time-averaged \swift-XRT spectrum covering the period 2021 February 7 to May 6 (MJD 59252--59340 inclusive) using the online \swift\ data products generator\footnote{\url{https://www.swift.ac.uk/user_objects/}} \citep{Evans09}.
The deadtime-corrected on-source time is 215~ks and the spectrum contains 114070 counts in the 0.3--10~keV range.

\subsubsection{Las Cumbres Observatory}
\label{sec:data-lco}
The Las Cumbres Observatory (LCO; \citealt{Brown13}) is a network of 25 small (0.4--2~m) telescopes located at seven sites around the globe.
\mcg\ has been monitored by the LCO with 4~day cadence since 2020 January, with the cadence intensifying to 1~day during the intensive \swift\ campaign.  
Sinistro cameras on the LCO 1-m telescopes were used on each visit to take a pair of CCD images per filter, enabling internal consistency checks for statistical errors and facilitating the identification of outliers due to cosmic ray hits or other anomalies. Exposure times were 120~seconds in the $u$ band, 30~seconds in the B, $g$, V, $r$, and $i$ bands, and 60~seconds in $z_s$, typically yielding photometric uncertainties of approximately 1\%, as estimated from the lightcurves of comparison stars.
The reduced CCD images were obtained from the LCO Archive\footnote{\url{http://archive.lco.global}}, having undergone standard bias subtraction and flat-field correction. 
Photometric measurements were extracted using the AGN Variability Archive (AVA) pipeline\footnote{\url{http://alymantara.com/ava}}. Detailed procedures for photometric extraction and absolute calibration are described in \citet{HS20}. 
In summary, aperture photometry was carried out using {\sc SExtractor} \citep{Bertin96}, with a 5\arcsec\ radius aperture chosen to balance signal-to-noise ratio (S/N) and seeing variations across the monitoring campaign.
Zeropoints for individual images were determined by cross-matching field stars with the APASS catalog \citep{Henden18} and PAN-STARRS \citep[for $z_s$][]{Flewelling20}. A bootstrapping simulation was employed to estimate the zeropoint and its associated $1\sigma$ uncertainty. These zeropoint corrections were applied to all lightcurves prior to the intercalibration described below.

\subsubsection{Dan Zowada Memorial Observatory}
\label{sec:data-zowada}
The Dan Zowada Memorial Observatory (hereafter Zowada: \citealt{Carr22}) is a fully-robotic 0.5~m telescope operated by Wayne State University and which, at the time of this campaign, was situated in New Mexico, USA.
It is capable of providing photometery in the Sloan bands $u^\prime$, $g^\prime$, $r^\prime$ and $i^\prime$ (\citealt{Fukugita96}) and the Pan-STARRS $z_\mathrm{s}$ band (\citealt{Tonry12}); hereafter we simply refer to these bands as $u$, $g$, $r$, $i$ and $z$.
Zowada monitored \mcg\ with approximately daily cadence between 2020 September and 2021 April, overlapping with the intensive \swift\ monitoring mentioned above.
The exposure times were 300 seconds, with 2 exposures per visit in $g$, $r$ and $i$, 3 in $z$ and 4 in $u$.

The data reduction pipeline is described in \cite{Carr22}.
Differential aperture photometry was performed following the procedure outlined in \cite{Miller23,Miller25}.  
Briefly, photometry was performed using a circular source aperture with radius 5 pixels, and a background annulus with inner and outer radius of 20 and 30 pixels.  
The total flux from three comparison stars was assumed to be constant over time in order to determine the lightcurve of the AGN in each band.
Given the high cadence of the observations from each facility during our campaign (Table~\ref{tab:obs}), we were able to flux-calibrate the Zowada lightcurves with respect to the LCO ones in each matching filter.  
This involved preforming a simple, linear interpolation onto a common series of times, then determining the flux offset and scaling factor to be applied to the (normalised) Zowada lightcurves in order to bring them into agreement with the LCO ones (in mJy).
Having then converted the Zowada lightcurves to mJy, the LCO and Zowada lightcurves were merged.
The resulting combined optical-infrared lightcurves are shown in Fig.~\ref{fig:optlcs}.

\begin{table*}
    \centering
    \caption{Summary of observations made during the \swift\ intensive monitoring period}
    \begin{tabular}{cc|cccc}
    \hline
    Observatory    & Filter & First obs.\        & Last obs.\          & $N_\mathrm{obs}$ & Cadence \\
                   &        & (MJD)              & (MJD)               &                  & [days]  \\
    \hline
    \swift\        & X-ray  & 2021 Feb 7 (59252) & 2021 May 6 (59340)  & 259              & 0.34 \\
    \swift\        & UVW2   & "                  & "                   & 218              & 0.41 \\
    \swift\        & UVM2   & "                  & "                   & 202              & 0.44 \\
    \swift\        & UVW1   & "                  & "                   & 192              & 0.46 \\
    \swift\        & U      & "                  & "                   & 194              & 0.45 \\
    \swift\        & B      & "                  & "                   & 216              & 0.41 \\
    \swift\        & V      & "                  & "                   & 238              & 0.37 \\
    LCO $+$ Zowada & $u$    & 2021 Feb 8 (59253) & 2021 Apr 21 (59325) & 83               & 0.87 \\
    LCO $+$ Zowada & $g$    & 2021 Feb 7 (59252) & 2021 May 2 (59336)  & 91               & 0.92 \\
    LCO $+$ Zowada & $r$    & "                  & 2021 May 1 (59335)  & 87               & 0.95 \\
    LCO $+$ Zowada & $i$    & "                  & "                   & 89               & 0.93 \\
    LCO $+$ Zowada & $z$    & "                  & 2021 Apr 23 (59327) & 83               & 0.90 \\
    \hline
    \end{tabular}
    \label{tab:obs}
\end{table*}

\subsection{Archival observations}
\subsubsection{\textit{NuSTAR}}
\label{sec:data-nustar}
A 25~ks hard X-ray observation of \mcg\ was made by the \textit{Nuclear Spectroscopic Telescope Array} (\nustar; \citealt{Harrison13}) on 2021 December 18 (observation ID: 90701640002), approximately seven months after the conclusion of our intensive \swift\ monitoring period.
These data were originally obtained for a study of the X-ray and mm intraday variability of \mcg\ by \cite{Petrucci23}.
We reduced these data with up-to-date software (\nustar\ Data Analysis Software v2.1.1) and calibration files (CALDB version 20210315).
The source spectra for each focal plane module (FPMA and FPMB) were extracted from circular regions of radius 60$^{\prime\prime}$ centred on the source.
The background spectra were extracted from 120$^{\prime\prime}$-radius regions away from the source.
The net 3--50~keV source count rates were $1.047\pm0.007$ and $0.967\pm0.006$ counts\,s$^{-1}$ in FPMA and FPMB, respectively. 
\textcolor{black}{We model these spectra in Section~\ref{sec:xspec}.}

\textcolor{black}{
\subsubsection{\textit{XMM-Newton}}
\label{sec:data-xmm}
A 38~ks X-ray observation of \mcg\ was made by the \xmm\ Observatory (\citealt{Jansen01}) on 2004 April 10 (observation ID: 0201930201). 
\cite{Matt06} reported $\approx10$~per cent X-ray flux variations and no spectral variations during the observation. 
The authors then modelled the time-averaged EPIC-pn spectrum with a power-law continuum plus Compton reflection model.
They found a relatively hard continuum spectrum, $\Gamma\approx1.8$, and substantial soft X-ray absorption by both neutral 
($N_\mathrm{H}=1.8\times10^{21}$\,cm$^{-2}$) and ionised ($N_\mathrm{H}=1.1\times10^{22}$\,cm$^{-2}$) gas within \mcg.
No excess soft X-ray emission above the power-law continuum was detected, and
relativistic broadening of the Fe\,K$\upalpha$ fluorescence line was ruled out.
We employ these archival data in Section~\ref{sec:xvar}.
}

\section{Timing analyses}\label{sec:timing}
As can be seen by eye in Figs.~\ref{fig:lcs} and \ref{fig:optlcs}, substantial flux variability can be seen in all of the lightcurves.
The lightcurves are clearly correlated, and the same variability patterns are apparent at all wavelengths. 
We therefore perform timing analyses to characterise the observed variability, assess the degree of correlation and determine the wavelength dependence of time lags between the different bands.

\textcolor{black}{These timing analyses will provide empirical links between the observed multi-band lightcurves. 
By quantifying the variability amplitudes, inter-band correlations and wavelength-dependent delays, 
we can later test whether the observed behaviour is consistent with a pure disc or BLR reprocessing (if either).
The resulting parameters form the basis for the physical interpretation developed in Sections~\ref{sec:sed} and \ref{sec:models}.}

\begin{figure*}
    \centering
    \includegraphics[width=0.9\textwidth]{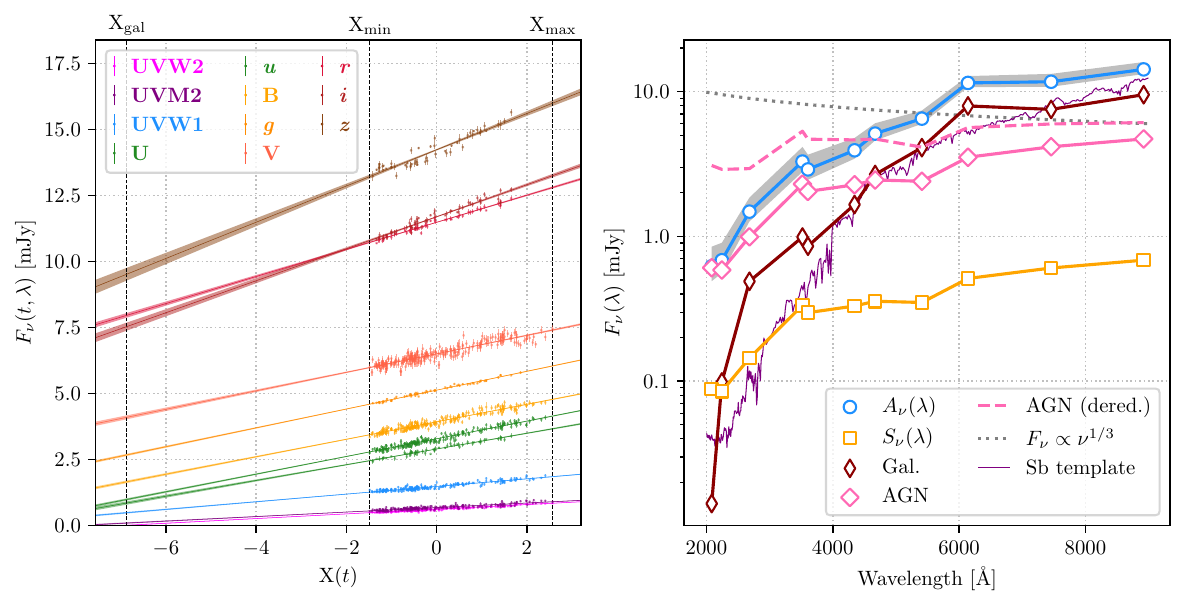}
    \caption{Flux-flux analysis of the \swift\ UVOT, LCO and Zowada data, used to separate the variable and constant components of the lightcurves.
    The left-hand panel shows the observed flux densities against a dimensionless lightcurve shape X($t$), which is normalised to mean 0 and rms 1.
    A linear regression extrapolates the flux-flux relations back to X$_\mathrm{gal}$, at which point the variable component is zero
    and the values of $F_\nu(\lambda,t)$ give the spectrum of the constant, host galaxy, component.  
    The right-hand panel shows the galaxy component as a the deep red line and open diamonds.
    (The thin, purple line shows the template spectrum of a barred spiral `Sb' galaxy, reddened for comparison with the other observed spectra.) 
    The right-hand panel also shows the average SED, $A_\nu(\lambda)$, i.e.\ $F_\nu(\lambda,t)$ where $\mathrm{X}(t)=0$, as a blue line with open circles.  
    The grey envelope, evaluating the linear model at  X$_{\rm max}$ and X$_{\rm min}$, shows the AGN flux range seen during the campaign.
    The difference between the average SED, $A_\nu(\lambda)$, and the galaxy component represents the average spectrum of the observed variable AGN component, shown as a solid pink line and pink diamonds. The dashed pink line shows the average AGN component corrected for Galactic reddening ($A_\mathrm{V}=0.568$~mag). 
    The rms of the variable AGN component, $S_\nu(\lambda)$, is shown as an orange line with open squares.
    The dotted grey line shows a canonical optical accretion disc spectrum with $F_\nu\propto\nu^{1/3}$ scaled to match the dereddened AGN spectrum in the IR.}
    \label{fig:fluxflux}
\end{figure*}

\subsection{Flux-flux analysis and fractional variability}
\label{sec:fluxflux}
\textcolor{black}{In this subsection we separate the variable AGN emission from the constant host-galaxy light so that we can more accurately determine the strength of the reverberation signal.}
\subsubsection{Flux-flux analysis and AGN / host galaxy decomposition}
In order to separate the variable and constant components within the lightcurves, we fit the multiband flux densities ($F_\nu[\lambda,t]$ in mJy) as a function of wavelength and time in the form  
\begin{equation}
    F_\nu(\lambda,t) = A_\nu(\lambda) + S_\nu(\lambda) \mathrm{X}(t)
\end{equation}
where $\mathrm{X}(t)$ is a dimensionless lightcurve scaled such that $\langle \mathrm{X} \rangle = 0$ and $\langle \mathrm{X}^2 \rangle = 1$; 
$A_\nu(\lambda)$ represents the mean spectrum, i.e.\  $F_\nu(\lambda,t)$ when $\mathrm{X}(t)=0$; and 
$S_\nu(\lambda)$ is the rms spectrum, i.e.\ the variable component.
An iterative decomposition procedure is used to determine $A_\nu(\lambda)$ and $S_\nu(\lambda)$, as described in Appendix~\ref{sec:fluxflux-details}. 

The results of this flux-flux analysis are shown in Fig.~\ref{fig:fluxflux}, and the values are given in Appendix~\ref{sec:fluxflux-details} (Table~\ref{tab:fluxflux}).
The linear flux-flux relations for each photometric band (the left-hand panel of Fig.~\ref{fig:fluxflux}) indicate that there is little or no colour change in the variable component as it brightens and fades. 
A simple model combining a variable flux component plus a constant flux component, both of fixed spectral shape, is therefore a reasonable representation of the observations.

The values of $F_\nu(\lambda,t)$ in the different wavebands, extrapolated back to where $F_\nu(\lambda,t)$ is one sigma above zero in the bluest waveband (UVW2), provide the SED of the constant, host galaxy, component. The difference between the mean overall spectrum, $A_\nu(\lambda)$, and the host galaxy gives the mean spectrum of the variable AGN component. These spectra are shown in the right-hand panel of Fig.~\ref{fig:fluxflux}.
We see that, as expected, the AGN component is bluer than the galaxy component.
The total spectrum is dominated by the host galaxy longward of the g-band ($\approx4500$~\AA) but by the AGN at shorter wavelengths.
In Fig.~\ref{fig:fluxflux} we also show the template spectrum of a barred spiral galaxy (appropriate for \mcg: \citealt{DEV91}) taken from the SWIRE template library \citep{Polletta07}.
Our flux-flux derived galaxy spectrum matches the template reasonably well in the optical-infrared but less so in the ultraviolet, in which the template is generally redder (although the differences are exaggerated by the logarithmic scale of the plot).
The bluest point (UVW2) of the galaxy spectrum lies below the template since, by construction, we assume the galaxy flux here is near zero (see Appendix~\ref{sec:fluxflux-details}).
The shape and scale of the blue end of the spectrum clearly depend on the choice of where to set X$_\mathrm{gal}$.
It is also possible that the flux-flux procedure does not produce a precisely correct spectral decomposition, a point we return to in Section~\ref{sec:discussion}.
Similar to what we see here, \cite{Prince25} also find that their flux-flux galaxy spectrum is a poorer fit in the ultraviolet to a template selected for NGC\,7469.

\textcolor{black}{Having estimated the constant host galaxy spectrum, we can now assess the variability of the AGN emission in isolation.}

\subsubsection{Fractional Variability}
We calculate the relative amount of variability in each band\textcolor{black}{, both in the raw and host galaxy subtracted lightcurves, informed by the results from our flux-flux analysis above}. 
The fractional variability $F_\mathrm{var}$ is defined in the standard way: given a set of fluxes $f$ with uncertainties $\sigma_f$ and variance $S^2$,
\begin{equation}
    F_\mathrm{var} = \sqrt{ \frac{S^2 - \langle \sigma_{f}^2 \rangle}{\langle f \rangle^2}}
\end{equation}
and the associated uncertainty is calculated following  \cite{Vaughan03}.
We calculate $F_\mathrm{var}$ for both the raw lightcurves and again after subtracting the (constant) host galaxy flux, as determined above.  
Both sets of $F_\mathrm{var}$ values are listed in Table~\ref{tab:fvar_lags}.
As expected, the fractional variability in X-rays ($\approx0.32$) is much greater than that in the UVOIR bands ($\lesssim0.15$).
When measuring $F_\mathrm{var}$ of the raw lightcurves, there is a pronounced drop in variability with increasing wavelength up to the $r$ band, beyond which we find the $i$ and $z$ bands have similar $F_\mathrm{var}$ to $r$ ($\approx0.04$).
This decrease in the fractional variability is likely due to the dilution by the (constant) host galaxy flux which has an increasing contribution at longer wavelengths. 
Subtracting the flux-flux model host galaxy fluxes obviously makes little difference in the UV bands, 
but results in fairly constant $F_\mathrm{var}$ across the optical and infrared bands ($\approx0.12$), with an excess in the $g$ and V bands around 5000~\AA.
Additionally, we note that in the six months preceding the intensive \swift\ monitoring period the mean, host galaxy subtracted fluxes in the optical-infrared bands were $\approx42$~per cent greater and\textcolor{black}{,} during this period\textcolor{black}{,} the fractional variability of the lightcurves decreases progressively with wavelength from $0.131\pm0.03$ in $u$ to $0.082\pm0.003$ in $z$.

\textcolor{black}{Overall, the flux–flux analysis confirms that the variable component is nearly colour-independent. 
After subtracting the host galaxy contribution to the lightcurves, the intrinsic AGN fractional variability remains roughly constant across the UVOIR range.}

\begin{figure*}
    \centering
    \includegraphics[width=1.70\columnwidth]{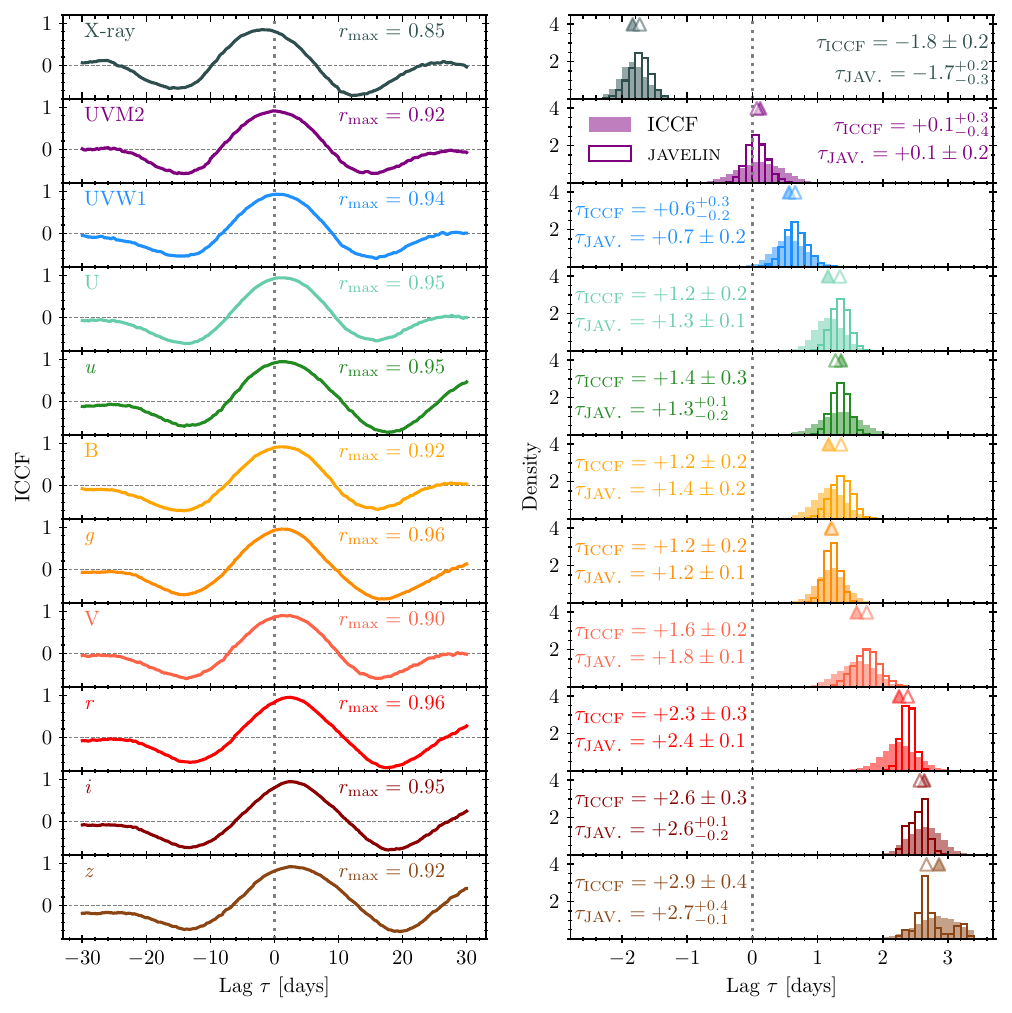}
    \caption{Results from the time series analyses of the \mcg\ intensive continuum reverberation mapping campaign.
    All measurements are relative to the \swift\ UVW2 band (2084~\AA).
    The left-hand panels show the interband interpolated cross-correlation functions (ICCFs), with their maxima $r_\mathrm{max}$ given in the inset text.
    The corresponding right-hand panels contain the cross-correlation centroid distributions (CCCDs), shown as solid histograms.
    Also shown are the \textsc{javelin}-derived lag distributions (open histograms).
    Filled and open triangles indicate the centroid lag $\tau$ for the ICCF and \javelin\ methods, respectively;
    the inset text quotes each lag and its uncertainty.
    }
    \label{fig:ccf_jav_results}
\end{figure*}

\subsection{ICCF interband lags}\label{sec:iccf}
\textcolor{black}{Having characterised the variability amplitudes, we next measure temporal relationships between the bands to determine any correlations and time delays between variations.}
We first calculate the cross-correlations and lags between our lightcurves using the interpolated cross-correlation function (ICCF) method described by \cite{Peterson98} (we use a Python implementation of that routine: PyCCF\footnote{\url{http://ascl.net/code/v/1868}}; \citealt{Sun18}).
The method creates many Monte Carlo simulations of the data; in each realisation a random subset of the data points are selected (`random subset selection': RSS, estimating the uncertainty on the results due to sampling) and each data point has its flux modified by random noise on the scale of the flux uncertainty (`flux randomisation': FR, estimating the uncertainty on the results due to flux errors).
The centroid (above 80~per cent of peak correlation) and peak of each simulated ICCF is recorded and the lag (and its uncertainty) is then calculated from the location (and width) of both the cross-correlation centroid distribution (CCCD) and the cross-correlation peak distribution (CCPD).
Generally, lag measurements from the CCCD are preferred \citep{Peterson98}.

\begin{figure*}
    \centering
    \includegraphics[width=0.9\textwidth]{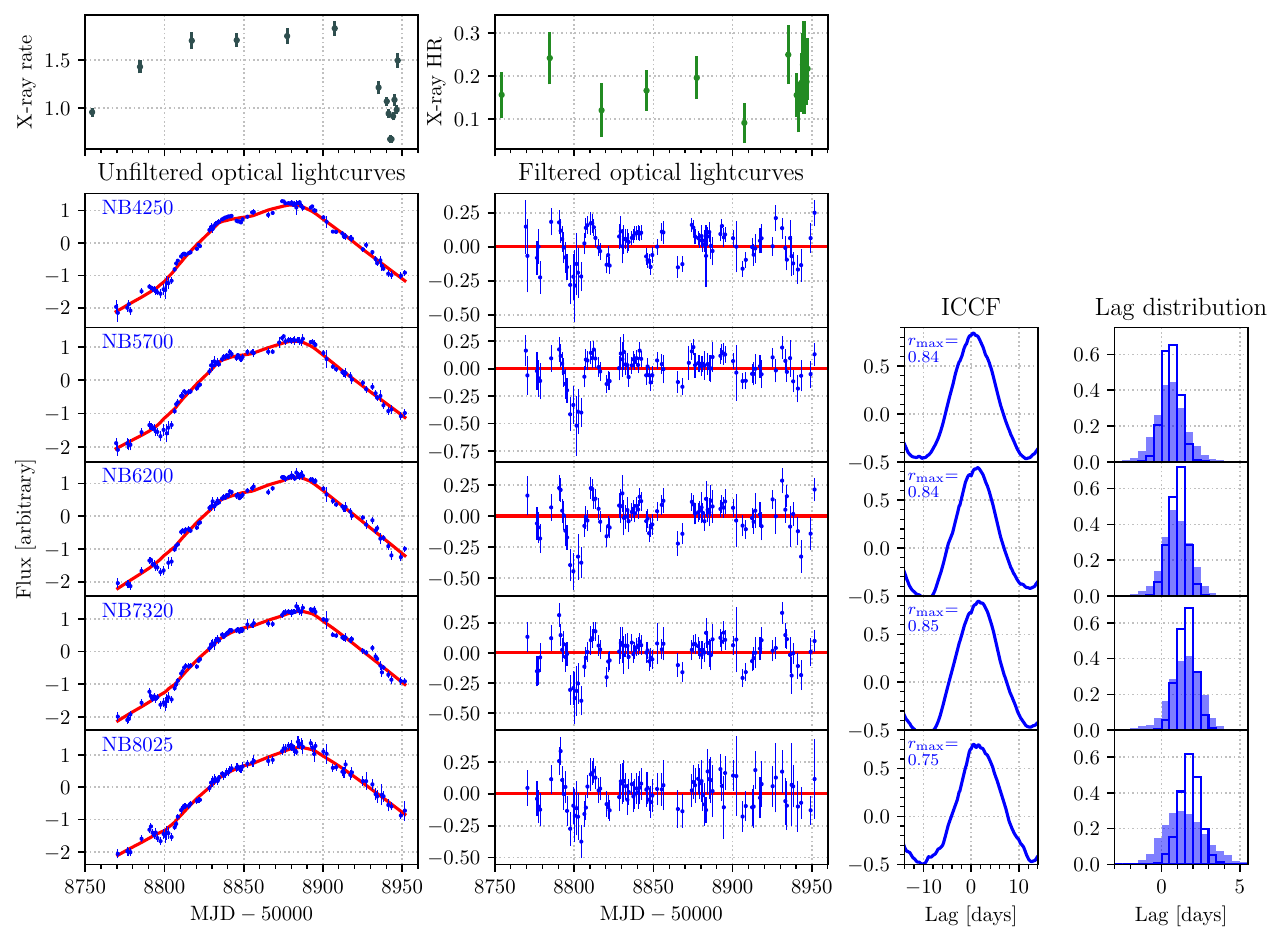} 
    \caption{\swift\ X-ray and Wise Observatory narrow band optical lightcurves of \mcg\ in the period 2019 October to 2020 April. 
    \textit{Top}:  The \swift\ XRT lightcurve (left) and hardness ratio (right; using 0.3--1.5 and 1.5--10~keV bands) are shown for comparison.
    \textit{Left}: The original lightcurves presented by \protect\cite{Fian23} in which long-term trend (shown in red) has been estimated using the LOWESS method with a 45~day smoothing window.
    (The optical fluxes have been normalised to have a mean of 0 and standard deviation of 1.)
    \textit{Centre}:  The filtered lightcurves, from which the long-term trend has been subtracted.
    \textit{Right:} The interpolated cross-correlation function (ICCF) and lag distribution for each filtered lightcurve relative to the 4250~\AA\ band.  Solid and open histograms show the ICCF centroid \textsc{javelin} lag distributions, respectively.}
    \label{fig:fian}
\end{figure*}

We used the  \swift\ UVW2 band as the reference lightcurve\footnote{This was the natural choice because it is the shortest-wavelength UVOIR lightcurve and has the highest S/N (Fig.~\ref{fig:lcs}).  It also makes comparisons with the \javelin\ analysis easier, since - because of the assumptions made in its modelling - \javelin\ requires the leading (or `driving') lightcurve to be input as the reference, and we expect variations at shorter wavelengths to lead those at longer wavelengths.} and, using an interpolation time step of 0.1~days and $10^4$ Monte Carlo iterations, we searched for lags in the range $\pm30$~days.
We find a strong correlation between the X-ray and UVW2 lightcurves ($r_\mathrm{max}=0.85$) and very strong correlations between UVW2 and the longer-wavelength lightcurves ($r_\mathrm{max}\geqslant0.90$).
Both the CCCD and the CCPD indicate an X-ray lag of $\approx-1.8$~days (i.e.\ variations in the X-ray lightcurve lead variations in the UV by $\approx1.8$~days).
Lags for the longer-wavelength bands are all positive (or are consistent with zero) and generally increase with wavelength up to $\approx3$~days for the $z$ band.
The full results are recorded in Table~\ref{tab:fvar_lags} and the ICCFs and CCCDs are shown in Fig.~\ref{fig:ccf_jav_results}.

\textcolor{black}{In summary, we measure strong correlations between all lightcurves and find a clear increase of lag with wavelength, with the X-rays leading the UV by $\approx1.8$~days and the \textit{z}-band by $\approx5$~days,  consistent with progressively larger reprocessing radii at longer wavelengths.}

\subsection{\javelin\ interband lags}\label{sec:jav}
\textcolor{black}{To verify the ICCF results and obtain statistically robust uncertainties on the lag measurements, we repeated the lag analysis using the model-based \javelin\ approach \citep{Zu16}.}
\javelin\ assumes that the AGN variability can be approximated by a damped, random walk \textcolor{black}{(DRW)}.
It creates a statistical model of the specfied driving lightcurve and uses this as a prior to model each lagged lightcurve, assuming that each is a smoothed, scaled, and time-shifted version of it.
In the following, we have used version 0.35 of \javelin\ in spectroscopic RM mode.
In this mode, \javelin\ fits for the scale and amplitude of the DRW model (determined from the driving lightcurve) and then for the lag, scale and the width of the smoothing function for each longer-wavelength lightcurve. 

To determine the X-ray lead, we input the X-ray lightcurve as the driving lightcurve for the UVW2 band.
For all other measurments, we have used the UVW2 lightcurve as the driving lightcurve, as was the case for the ICCF analysis.
Informed by the ICCF analysis (Section~\ref{sec:iccf}), we searched for lags in a narrower range ($\pm5$~days).
Since we expect the smoothing to be on the order of the lag, we also restricted the width of the smoothing top-hat function to $<5$~days.

The \javelin\ lags are reported in Table~\ref{tab:fvar_lags}, and the lag distributions derived from its MCMC sampling are shown as open histograms in Fig.~\ref{fig:ccf_jav_results}. 
The \javelin\ lags are in very good agreement with the ICCF ones, again revealing an X-ray lead of $\approx1.7$~days and lags relative to UVW2 increasing with wavelength up to $\approx3$~days in the near-infrared $z$ band. 
In the UVOIR bands, the \javelin\ lag uncertainties are smaller than the corresponding ones determined by the ICCF method, which can be seen by eye in Fig.~\ref{fig:ccf_jav_results}.
These smaller uncertainties are expected, since tests performed on simulated lightcurves have demonstrated that \javelin\ produces lower and more accurate error estimates than the ICCF method (e.g.\ \citealt{Li19}, \citealt{Yu20}), provided that the simplest underlying assumptions of reverberation are valid.
Additionally, since we adopted the RSS procedure in calculating the ICCF lags (which is not essential for well-sampled lightcurves such as ours) the associated uncertainties will be increased as a result of not using the full data sets.

\textcolor{black}{The close agreement between \javelin\ and ICCF lags reinforces that the observed correlations relate to genuine reverberation delays rather than artefacts of the sampling.
We discuss this further in Appendix~\ref{app:javelin}.}

\begin{table*}
    \centering
    \caption{X-ray, UV, optical and infrared fractional variabilities, and inter-band correlations and observed-frame lags.}
    \begin{tabular}{cc|cccccc}
    \hline
    (1)    & (2)                      & (3)              & (4)              & (5)                     & (6)                  & (7)                  & (8) \\
    Filter & $\lambda_\mathrm{eff}$   & $F_\mathrm{var}$ & $F_\mathrm{var}$ & $r_\mathrm{max}$        & CCCD lag             & CCPD lag             & \javelin\ lag \\
           & [\AA]                    &                  & gal.\ sub.\      &                         & [days]               & [days]               & [days] \\
    \hline
    X-ray & --                        & $0.317\pm0.004$  & --               & 0.85                    & $-1.8\pm0.2$         & $-1.8^{+0.8}_{-0.2}$ & $-1.7^{+0.2}_{-0.3}$  \\ 
    UVW2  & 2083.95                   & $0.142\pm0.002$  & $0.143\pm0.002$  & --                      & --                   & --                   & --                    \\
    UVM2  & 2245.03                   & $0.127\pm0.004$  & $0.131\pm0.004$  & 0.92                    & $+0.1^{+0.3}_{-0.4}$ & $-0.1\pm0.4$         & $+0.1\pm0.2$          \\
    UVW1  & 2681.67                   & $0.094\pm0.003$  & $0.102\pm0.003$  & 0.94                    & $+0.6^{+0.3}_{-0.2}$ & $+0.5^{+0.6}_{-0.7}$ & $+0.7\pm0.2$          \\
    U     & 3520.88                   & $0.100\pm0.002$  & $0.114\pm0.002$  & 0.95                    & $+1.2\pm0.2$         & $+1.1^{+0.4}_{-0.5}$ & $+1.3\pm0.1$          \\
    $u$     & 3608.04                   & $0.082\pm0.003$  & $0.114\pm0.005$  & 0.95                    & $+1.4\pm0.3$         & $+1.3^{+0.4}_{-0.7}$ & $+1.3^{+0.1}_{-0.2}$  \\
    B     & 4345.28                   & $0.082\pm0.001$  & $0.112\pm0.002$  & 0.92                    & $+1.2\pm0.2$         & $+1.2\pm0.5$         & $+1.4\pm0.2$          \\
    $g$     & 4671.78                   & $0.060\pm0.001$  & $0.127\pm0.002$  & 0.96                    & $+1.2\pm0.2$         & $+1.2^{+0.4}_{-0.3}$ & $+1.2\pm0.1$          \\
    V     & 5411.45                   & $0.054\pm0.001$  & $0.139\pm0.004$  & 0.90                    & $+1.6\pm0.2$         & $+1.5^{+0.8}_{-0.6}$ & $+1.8\pm0.2$          \\
    $r$     & 6141.12                   & $0.037\pm0.001$  & $0.121\pm0.004$  & 0.96                    & $+2.3\pm0.3$         & $+2.3^{+0.3}_{-0.4}$ & $+2.4\pm0.1$          \\
    $i$     & 7457.89                   & $0.043\pm0.001$  & $0.119\pm0.004$  & 0.95                    & $+2.6\pm0.3$         & $+2.5\pm0.3$         & $+2.6^{+0.1}_{-0.2}$  \\
    $z$     & 8922.78                   & $0.041\pm0.001$  & $0.118\pm0.006$  & 0.92                    & $+2.9\pm0.4$         & $+2.6^{+0.4}_{-0.6}$ & $+2.7^{+0.4}_{-0.1}$  \\
    \hline
    \end{tabular}
    \parbox{12.5cm}{All values relate to the three-month period of intensive \swift\ monitoring (2021 February--May).
    The correlation coefficient and lags are measured with respect to the \swift\ UVW2 band.
    \textit{Columns:} (1) photometric filter; (2) effective wavelength of the photometric filter; (3) fractional variability of the raw lightcurve; (3) fractional variability of the host galaxy flux subtracted lightcurve; (5) maximum correlation coefficient; (6) cross-correlation centroid distribution lag; (7) cross-correlation peak distribution lag; (8) the lag determined by \javelin.
        }
    \label{tab:fvar_lags}
\end{table*}

\subsection{Reanalysis of the 2019-20 Wise Observatory lightcurves}
\label{sec:fian}
\begin{figure*}
    \centering
    \includegraphics[width=16cm,angle=270]{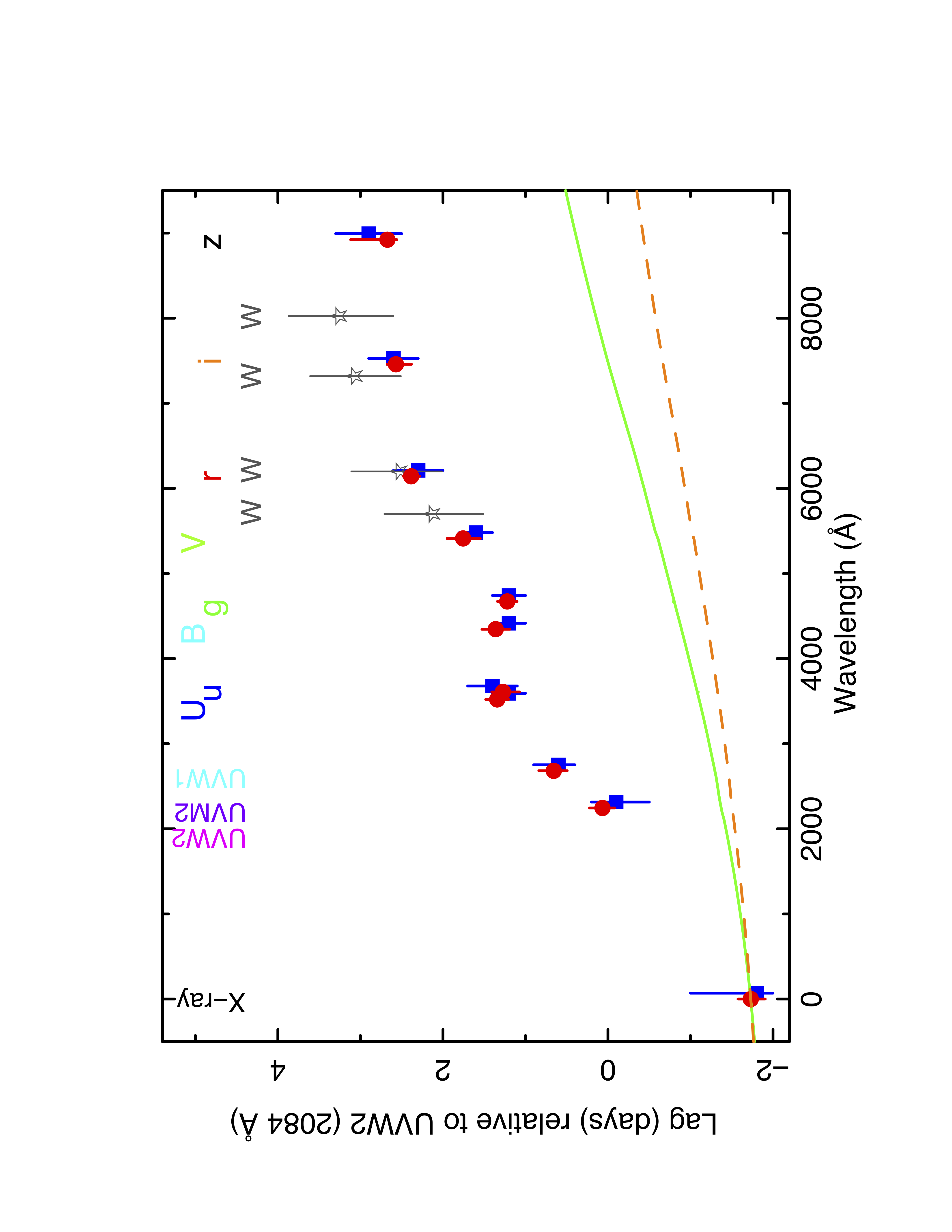}
   \vspace*{-10mm}
   \caption{The lag spectrum of \mcg\ from the intensive \swift\ and LCO monitoring in 2021. All lags are measured relative to the \swift\ UVW2 band (2084~\AA). The red filled circles are lags measured using \javelin\ and the blue filled squares are the ICCF centroid lags. The lag distributions for the individual bands are shown in Fig.~\ref{fig:ccf_jav_results}. The solid green line is a simple numerical model fit to the lags expected from X-ray reprocessing from a disc around a non-rotating black hole of mass $2.82\times 10^{7}$~M$_\odot$ and accretion rate of 20 per cent of Eddington. The dashed orange line is for reprocessing from a disc around a maximally rotating Kerr black hole. The open grey stars are the lags derived from the observations by \protect\cite{Fian23} at the Wise Observatory after filtering out long-term trends in the lightcurves.}
    \label{fig:lag_spec}
\end{figure*}

\textcolor{black}{
In the preceding subsections we determined the X-ray to IR lag spectrum from the lightcurves obtained during our intensive \swift\ monitoring campaign.
Because a prior campaign reported much longer optical lags, we revisit those data to test whether the differences reflect sampling effects or physical changes.}
The lags we have determined for the intensive \swift\ monitoring period are much shorter than those recently reported by \cite{Fian23} following an optical photometric monitoring campaign on \mcg\ conducted between 2019 October and 2020 April using the C18 telescope at the Wise Observatory in Israel.
These observations span a period between 16 and 10 months prior to the beginning of our intensive \swift\ monitoring.
The Wise-C18 observations were made with approximately nightly cadence in five narrow optical bands (centred on 4250, 5700, 6200, 7320 and 8250~\AA), chosen to sample the optical AGN continuum emission and avoiding broad emission lines.
Fig.~\ref{fig:fian} shows that the optical variability in the earlier campaign is different in character to that seen during our \swift\ monitoring period. 
The Wise lightcurves show mainly a slow ($\sim$months), smooth rise and subsequent fall in brightness.
In contrast our \swift\ lightcurves (Fig.~\ref{fig:lcs}) exhibit more prominent, faster variability (over $\sim$days--weeks).
We also note that the mean optical flux during the earlier campaign is approximately a factor of four higher than that we observed: \cite{Fian23} report $\langle{F_{4250\,\si\angstrom}}\rangle=8.48$~mJy, compared with $\langle{F_\mathrm{B}}\rangle=2.27$~mJy during the intensive \swift\ monitoring period\footnote{Both estimates of the flux density have had the host galaxy contribution subtracted.  \cite{Fian23} also note that their optical fluxes are also a factor two higher than those reported by \cite{Fausnaugh17}.}.
\cite{Fian23} employed several different methods to determine the optical interband lags and found that the mean lag of the 8250~\AA\ band relative to 4250~\AA\ was $7.1\pm1.1$~days.  From Fig.~\ref{fig:lag_spec} it can be seen that the lag between these wavelengths is only $\approx1.6$~days in our data.

We have reanalysed the optical lightcurves of \cite{Fian23} using the ICCF method and \javelin\ and confirm their lag determinations and very steep lag spectrum. In some previous monitoring observations it has been found that the presence of short lags can be obscured if there are also longer term trends in the lightcurves \citep{Welsh99, McHardy14, Pahari20}. Removal of such trends can reveal the shorter lags. 
Here we fitted the long-term trends in the lightcurves using a Locally Weighted Scatterplot Smoothing (LOWESS) filter with a width of $\approx45$~days.
This fit results in smoothed optical lightcurves that contain the variability on $\approx$month timescales.
Subtracting the smoothed lightcurves from the observed ones leaves behind just the faster variability (Fig.~\ref{fig:fian}).
We then determined the lags between these filtered lightcurves using both the ICCF method and \javelin.
The results of these tests are contained in Table~\ref{tab:fian}.
The lags between the filtered lightcurves are much shorter than those reported by \cite{Fian23} and (as shown in Fig.~\ref{fig:lag_spec}) are entirely consistent with our own lags from the 2021 data set.

\textcolor{black}{In Appendix~\ref{app:detrend} we further investigate the effects of detrending the lightcuves on the recovered lags.
We conclude that the shorter lags determined following the detrending procedure do not result simply from the removal of the low-frequency components of a single, broad response.
Rather, it appears that the filtering distinguishes two separate components in the signal: 
a fast, reprocessing component and slower, large-amplitude trend of unknown origin 
(possibly resulting from slow-moving temperature fluctuations across the disc, e.g.\ \citealt{Neustadt22}).
In summary, the earlier long lags were largely the result of slow, large-amplitude trends in the lightcurves; 
shorter lags, consistent with our own, are retrievable in the earlier data following detrending.
Having verified our lag measurements we next assemble a full lag spectrum and conduct a first pass at comparing it to a physical model.}

\begin{table}
    \centering
    \caption{Optical band lags for the 2019-20 Wise Observatory data}
    \begin{tabular}{l|ccccc}
    \hline
           & \multicolumn{3}{c}{Unfiltered} & \multicolumn{2}{c}{Filtered} \\
           & \multicolumn{3}{c}{$\overbrace{\rule{4cm}{0pt}}$} & \multicolumn{2}{c}{$\overbrace{\rule{2.6cm}{0pt}}$} \\
    Filter & ICCF                 & \textsc{javelin} & \textsc{javelin}    & ICCF                   & \textsc{javelin} \\
           & (centroid)           & (spectro.)       & (photo.)            & (centroid)             & (spectro.)       \\
    \hline
    NB5700 &  $1.0^{+1.1}_{-1.0}$ & $1.3\pm0.3$      & $2.2^{+1.6}_{-0.8}$ & $0.55^{+0.98}_{-0.88}$ & $0.77^{+0.58}_{-0.62}$ \\
    NB6200 &  $2.1^{+1.0}_{-1.1}$ & $2.7\pm0.3$      & $2.8^{+0.8}_{-0.7}$ & $0.96^{+0.83}_{-0.81}$ & $1.17^{+0.58}_{-0.52}$ \\
    NB7320 &  $5.1^{+1.3}_{-0.8}$ & $3.8\pm0.4$      & $4.1^{+0.7}_{-0.6}$ & $1.54^{+0.90}_{-0.98}$ & $1.71^{+0.54}_{-0.56}$ \\
    NB8025 &  $7.4^{+1.4}_{-1.1}$ & $5.3\pm0.6$      & $7.3^{+1.5}_{-1.2}$ & $1.44^{+1.41}_{-1.29}$ & $1.90^{+0.61}_{-0.66}$ \\
    \hline
    \end{tabular}
    \parbox{\columnwidth}{\textit{Notes:} All lags are measured relative to the narrow-band 4250~\AA\ filter.
    The lags for the `unfiltered' lightcurves were measured from the original data and were reported by \protect\cite{Fian23}.
    The `filtered' lags have been determined from the modified lightcurves, after removal of the long-term ($\sim1$~month) trends (see Section~\ref{sec:fian} in the text for details).
    }
    \label{tab:fian}
\end{table}

\subsection{Combined Lags and Preliminary Modelling}
\label{subsec:initial_mod}
\textcolor{black}{We now combine all lag measurements from the above to construct a complete lag-versus-wavelength spectrum and make a first comparison with a simple, standard accretion disc model.
This will inform us how well our data conform to basic assumptions about the reverberation process and whether more detailed modelling is required.}
In Fig.~\ref{fig:lag_spec} we present all the observed lags, relative to the \swift\ UVW2 band, derived using both ICCF (blue squares) and \javelin\ (red circles) for all of the \swift, LCO, and Zowada lightcurves from the season of intensive observation. We also present the \javelin\ lags from the de-trended Wise Observatory observations, with their zero point being adjusted so that the 4250~\si\angstrom\,-band has a lag of 1.36~days, which is the \javelin\ lag of the \swift\ B-band (4345~\si\angstrom) relative to UVW2. 
\textcolor{black}{Because these bands are close in wavelength, further interpolation would alter the lag negligibly, and given the uncertainties around that wavelength, no further adjustment is warranted. 
These Wise lags are shown only for comparison and are not used in the subsequent modelling.}

In Fig.~\ref{fig:lag_spec} we show model lags calculated using the same numerical code used by \cite{McHardy18} (based on illumination of an accretion disc with a temperature profile as described by \citealt{SS73}) who compared the observed and model lags of 5 AGN. Here we take a mass of $2.82 \times 10^{7}$~\msun, an accretion rate of 20~per cent of Eddington (based on \textsc{cream} modelling of the optical lightcurves by \citealt{Fausnaugh18}), an illuminating luminosity of $2.7 \times 10^{44}$~erg\,s$^{-1}$, based on a factor of 2 extrapolation of the \swift\ BAT 14--195~keV observed luminosity \citep{Lien25}, an albedo of 0.2 and a lamppost source model with a height of 10~\rg. 
\textcolor{black}{These parameters are similar to those used by \cite{McHardy18}; small changes in source height or illuminating luminosity have little effect on the UV--optical lags.}
\textcolor{black}{Neither that work nor the present model includes a colour-correction factor, which can influence the lag spectrum, although \citet{McHardy23} found that NGC\,4395 required almost none.} 
We \textcolor{black}{return} to this point in more detailed modelling.

\textcolor{black}{Predicted lags for a non-rotating Schwarzschild black hole (solid green line) and a maximally rotating Kerr black hole (dashed orange line) are shown in Fig.~\ref{fig:lag_spec}. 
They are calculated relative to the X-rays, rather than UVW2 as our data; however, for a source height of 10~\rg, the resulting offset of 0.03~days is negligible.}

\cite{McHardy18} found that the ratio between the observed and model lags, between the UV and V bands, was between a factor of 1 and 2 for a non-rotating black hole. Looking broadly at the overall shape of the lag spectrum we find the same result here with a ratio of $\sim2.5$. However, and again in agreement with \cite{McHardy18}, we find a larger discrepancy between the observed and model lags between the X-ray and UV bands.

In detail, however, we see that the observed lag spectra do not agree well with this simple model. Besides the discrepancy between the amplitude of the observed and model X-ray to UV lags, the curvature of the lag spectrum does not agree with the upwardly curving shape with increasing wavelength of $\tau \propto \lambda^{4/3}$ expected from simple disc reprocessing. In fact the gradient of the lag spectrum decreases with increasing wavelength, with the lags approaching an asymptotic value. It is actually \textcolor{black}{well described ($\chi^{2}=6.7, \mathrm{d.o.f.}=8, P = 57$~per cent using the ICCF centroid lags)} by $\tau \propto \lambda^{0.5}$ over all observed wavebands. Although not exactly the same, a similar approach of the lags to an asymptotic value at long wavelength was seen in NGC\,4395 \cite{McHardy23} and interpreted as an indication that there was an edge to the reprocessing region. In fact a decrease in the gradient of the model lag spectra can be seen at the very longest wavelengths as the model contains a disc edge at 4800~\rg\, at the dust sublimation temperature (here, 2000~K) on the surface of the disc.

However the most prominent features of the observed lag spectrum, which are not expected from disc reprocessing, are the two large bumps, where the lags are in excess of the lags in the surrounding bands, one centred around the $u$-band and the other near the $i$-band. These bumps are indicative of the Balmer and Paschen continua respectively, expected from reprocessing in the BLR \citep{KG01, KG19, Netzer20}. These bumps have been seen previously \citep[e.g.][]{Cackett18, Netzer22} but the present examples are perhaps the most prominent and clearest yet presented.

\textcolor{black}{In summary, the combined lag spectrum follows a form roughly $\tau\propto\lambda^{0.5}$, flatter than the canonical $\tau\propto\lambda^{4/3}$.
Additional complexity includes the excesses around the \textit{u} and \textit{r}--\textit{i} bands that may be indicative of reprocessing of the driving continuum by the BLR.
These features motivate more detailed, physically informed modelling.} 
In Section~\ref{sec:models} we will examine the origin of this complex lag spectrum in more detail. 
In order to calculate the most accurate models, we require the best available estimates of physical parameters such as the Eddington accretion ratio (\me) and the \textcolor{black}{SED} of the ionising radiation, which we discuss in Section~\ref{sec:sed}.

\section{SED shape and Determination of {\me} and mass}
\label{sec:sed}
\textcolor{black}{Having established the temporal behaviour in the preceding Section, we now use}
our time-averaged and archival multiwavelength data \textcolor{black}{to} analyse the spectra and broadband SED of \mcg\ to determine important physical parameters, particularly its Eddington accretion ratio (\me)\textcolor{black}{, and mass, parameters which determine both the geometry and therefore the lag spectrum of the source}.

\subsection{The hard X-ray spectrum and \me}\label{sec:xspec}
\begin{figure}
    \centering
    \includegraphics[width=\columnwidth]{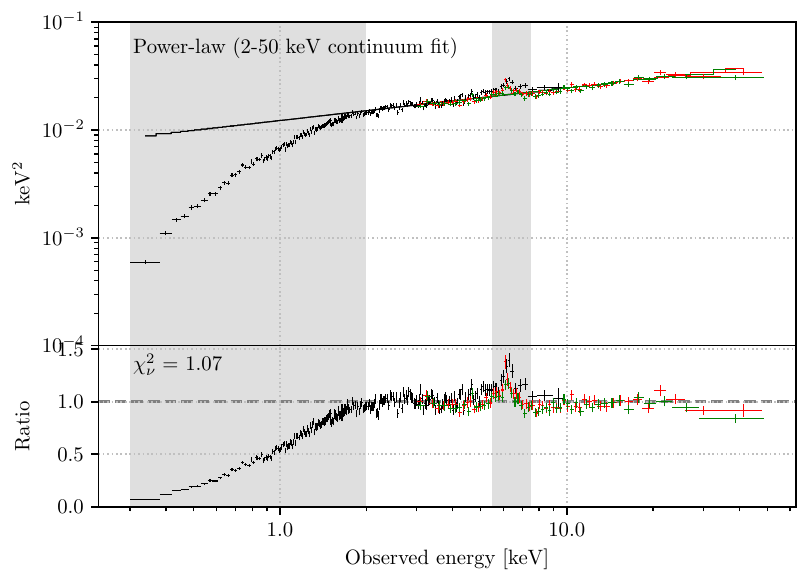}
    \caption{A power-law fit to the time-averaged \swift-XRT (black) and \nustar\ FPMA (red) and FPMB (green) spectra.
    Data in the grey regions were excluded from the fit.
    The upper panel shows the unfolded spectrum and model.  The lower panel shows the ratio of the data to the model.}
    \label{fig:xraypo}
\end{figure}

The hard X-ray spectral index is an indicator of \me\ \cite[e.g.][]{Kubota18}, with harder spectra being associated with lower accretion ratios. 
Here, we fit the time-averaged \swift-XRT spectrum from our campaign, together with the archival \nustar\ spectra recorded in 2021 December. 
\textcolor{black}{T}o \textcolor{black}{minimise} the complications \textcolor{black}{from} absorption and emission line features seen in previous observations, we \textcolor{black}{fit only the 2--5.5 and 7.5--50~keV data with a}
simple power-law \textcolor{black}{for which} we obtain \textcolor{black}{a photon index $\Gamma=1.697\pm0.008$ and} $\chi^2_\nu=1206/1121~\mathrm{d.o.f}=1.08$ (Fig.~\ref{fig:xraypo}).
The cross-normalisation factors applied to the \nustar\ FPMA and FPMB spectra are similar at $0.96\pm0.01$ and $0.94\pm0.01$, respectively. 
\textcolor{black}{W}e calculate the flux to be $\log(F_{2-10\mathrm{keV}}/\mathrm{erg\,s^{-1}\,cm^{-2}})=-10.302\pm0.003$\textcolor{black}{,} correspond\textcolor{black}{ing} to 
$L_{2-10\mathrm{keV}}=4.79\times10^{43}~\mathrm{erg\,s^{-1}}\approx1.3$~per cent of the Eddington luminosity, assuming \textcolor{black}{a} mass $2.82\times10^7~\mathrm{M}_\odot$ \citep{Fausnaugh17}.
\textcolor{black}{Using the luminosity-dependent X-ray bolometric correction of \citet{Duras20}, we estimate $\dot{m}_\mathrm{E} = L_\mathrm{bol}/L_\mathrm{Edd} \approx 0.23$.  
We return to this in Section~\ref{sec:bbsed} and Appendix~\ref{app:medd}, where other multiwavelength indicators of \me\ are compared.} 

\subsection{X-ray Variability Mass Estimates}\label{sec:xvar}
\textcolor{black}{We next use the short-timescale X-ray variability properties of \mcg\ to obtain an estimate of the black hole mass independent of optical reverberation-based methods.}
Black hole masses can be estimated from X-ray power spectral density (PSD) parameters, either from a bend in the PSD or from the normalisation of the high frequency part of the PSD \citep{McHardy88, McHardy13}. 
\textcolor{black}{We compute the excess variance of the 38~ks} \textit{XMM-Newton} observation of \cite{Matt06}, \textcolor{black}{binned at 250~s following \cite{Ponti12}, as}
0.218 (counts\,s$^{-1})^{2}$ and a mean count rate of 15.09 counts\,s$^{-1}$, giving a \textcolor{black}{normalised} excess variance $\sigma^{2}_\mathrm{rms}=9.57 \times 10^{-4}$.
\textcolor{black}{Using the 40~ks $\sigma^{2}_\mathrm{rms}$-$M_\mathrm{BH}$ relation of \cite{Ponti12} we obtain $M_\mathrm{BH}=5.9^{+3.5}_{-1.8} \times 10^{7}~\mathrm{M}_\odot$
(here the uncertainties in the mass are derived from the scatter in the relationship).}
\textcolor{black}{This is} more than twice the \textcolor{black}{optical RM} mass reported by \cite{Fausnaugh17} but is consistent with that lower mass within the uncertainties.

\textcolor{black}{We also combined the archival \xmm\ lightcurve with that of our \swift\ campaign to construct a PSD spanning $\log(\nu/\mathrm{Hz})=-7$ to $-3$, which we fit with a bending power-law using \psresp\ \citep{Uttley02}.
Since we expect the slope below the PSD bend to be poorly determined by our observations we fix that slope to $-0.8$, similar to that found in better sampled observations \cite[e.g.][]{McHardy04}.
The high frequency slope is well-constrained at $-2.0\pm0.5$ (90~per cent confidence bounds), with no measurable bend at low frequencies.
In an attempt to better localise the bend frequency we fixed the high frequency slope at $-2.0$ and refit the PSD.
The resulting fit is very good ($P=0.93$) although no bend was found in the observed frequency range ($>10^{-7}$~Hz).}
We searched for a bend in the range  $10^{-9}$ to $10^{-5}$~Hz as a bend below the observed frequencies will have a flattening effect on the PSD in the lowest observable frequencies. 
The best-fitting bend frequency was almost at the limit of the frequencies searched ($2.2 \times 10^{-9}$ Hz) although the 90 per cent confidence region \textcolor{black}{extends up to $1.5 \times 10^{-6}$~Hz}.
\textcolor{black}{We adopt the bolometric luminosity derived from $L_{5100\,\si\angstrom}$ by \cite{Fausnaugh17} and apply the \cite{McHardy06} relation between PSD bend frequency, luminosity and mass, finding $M_\mathrm{BH}\sim5$--$70\times10^7$~\msun.
(The lower end of this mass range relates to the upper uncertainty bound on the bend frequency and is consistent with the mass derived from the normalised excess variance.)}
We conclude that although the present X-ray data is not good enough to provide a \textcolor{black}{precise} measurement of the black hole mass, there are indications that the mass may be towards the upper limit of the range quoted by \cite{Fausnaugh17}\textcolor{black}{, i.e.} $\approx8.3\times10^7~\mathrm{M}_\odot$.

\textcolor{black}{Having now assessed both the spectral and timing properties of the X-ray data, we turn our attention to the UVOIR spectrum and overall SED to further refine our estimates of mass and \me.}

\subsection{The UVOIR continuum: reddening and \me}\label{sec:reddening}
\textcolor{black}{We next characterise the shape of the AGN UVOIR emission to estimate the intrinsic reddening within \mcg\ which, in turn, affects our estimate of $\dot{m}_\mathrm{E}$.}

\begin{figure}
    \centering
    \includegraphics[width=\columnwidth]{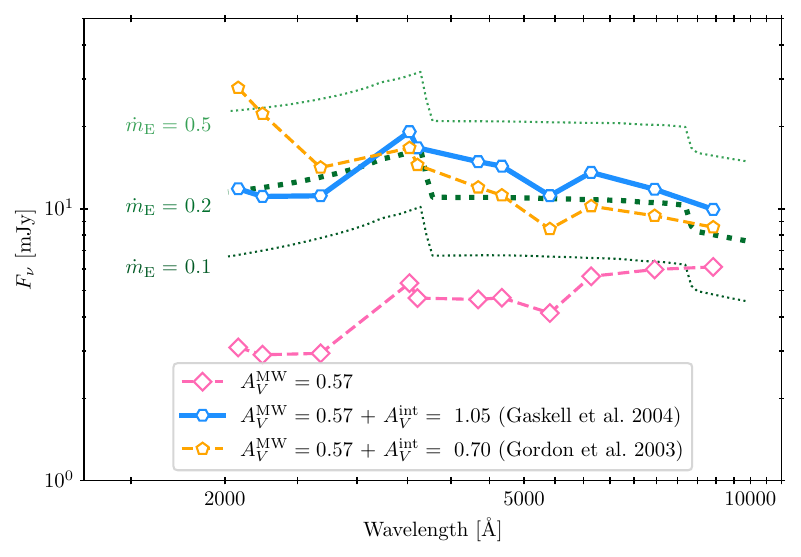} \\
    \caption{\textcolor{black}{The green dotted lines show accretion disc $+$ diffuse BLR continuum spectra for Eddington ratios $\dot{m}_\mathrm{E}=0.1$, 0.2 and 0.5.
    The disc spectra are predicted by the model \textsc{qsosed} and the diffuse BLR continua are scaled to these disc spectra according to \protect\cite{KG19}.}
    In pink diamonds we show the mean AGN spectrum from our flux-flux analysis, corrected \textcolor{black}{only} for Milky Way reddening $A_V^\mathrm{MW}=0.57$~mag.
    Orange pentagons show the spectrum additionally corrected for intrinsic reddening with $A_V^\mathrm{int}=0.70$~mag using the SMC Bar extinction curve of \protect\cite{Gordon03}.
    Blue hexagons show the spectrum additionally corrected for intrinsic reddening with $A_V^\mathrm{int}=1.05$~mag using the AGN extinction curve of \protect\cite{Gaskell04}\textcolor{black}{: this approximately matches the disc$+$BLR spectrum for $\dot{m}_\mathrm{E}=0.2$.}
}
    \label{fig:uvopt_spec}
\end{figure}

Fig.~\ref{fig:uvopt_spec} \textcolor{black}{shows in pink} the AGN \textcolor{black}{UVOIR} spectrum output by the flux-flux analysis (Section~\ref{sec:fluxflux}) corrected only for Galactic reddening assuming the extinction curve of \cite{CCM89} with $A_V=0.568$~mag \citep{SF11}.
The observed spectrum is much redder than \textcolor{black}{the expected} $F_\nu\propto\nu^{1/3}$ \textcolor{black}{suggesting either a cool disc or substantial intrinsic reddening.}  
\textcolor{black}{Strong X-ray extinction (e.g.\ \citealt{Matt06}; \citealt{Bianchi10}; \citealt{Patrick12}) and UV resonance-line absorption are evidence of a substantial amount of gas within \mcg\ along our line of sight to the nucleus, so dust may also be present.
\cite{Jaffarian20} estimated $E(B-V)=0.65$~mag from the optical Balmer decrement, further motivating our test of reddening here.}

\textcolor{black}{We compare the empirical UVOIR spectrum with model disc spectra for $\dot{m}_\mathrm{E}=0.1$, 0.2, and 0.5 from \textsc{qsosed} \citep{Kubota18}, assuming the \cite{Fausnaugh17} mass of $2.82\times10^7$~\msun\ to which we have added a diffuse BLR component, appropriately scaled following \cite{KG19}.
Rather than performing a formal fit, this was a simple heuristic exercise intended to identify plausible combinations of the intrinsic reddening and accretion rate.
Dereddening the observed spectrum using two extinction curves (a `flat' AGN curve from \citealt{Gaskell04} and the `steep' SMC-Bar curve of \citealt{Gordon03}), we find that the flat curve provides a reasonable match to the $\dot{m}_\mathrm{E}=0.2$ disc$+$BLR model for $A_V^\mathrm{int}=1.05$~mag, while the SMC curve over-corrects the UV flux (Fig.~\ref{fig:uvopt_spec}).
Assuming the \cite{Gaskell04} curve is appropriate, the dereddened UVOIR spectrum is consistent with \me$\approx0.20$, as implied by the X-ray spectrum.}
The flux correction factors for this curve and $A_V^\mathrm{int}$ are given in Table~\ref{tab:fluxflux}.

\textcolor{black}{Next, in Section~\ref{sec:bbsed}, we combine the UVOIR and X-ray spectra to produce an overall SED from which we make our best estimates of the mass, \me\ and reddening.}

\subsection{Broadband SED modelling: Mass, \me, and reddening}
\label{sec:bbsed}
With estimates of the intrinsic UVOIR and X-ray spectra in hand, we next attempt to obtain a self-consistent fit to the whole broadband SED using the models \textsc{qsosed} and \textsc{agnsed} \citep{Kubota18}. \textcolor{black}{Our aim is to determine the combination of mass, \me, and reddening which best explain the overall SED.}

We begin by assuming a non-spinning black hole with the mass fixed to the value $2.82\times10^7~\mathrm{M}_\odot$ given by \cite{Fausnaugh17}.  
\textcolor{black}{\textsc{qsosed} assumes a stratified accretion flow geometry consisting of (i) a spherical, hot corona around the black hole that produces hard X-rays; (ii) a warm, Comptonising inner disc that produces the soft X-ray excess emission; and (iii) a standard, cold, outer accretion disc from which the UVOIR spectrum arises (see figure 2 of \citealt{Kubota18}). In \textsc{qsosed} some parameters, such as the accretion rate and X-ray spectral index, are physically linked. \textsc{agnsed} assumes the same underlying geometry but is more flexible and some of the assumptions of \textsc{qsosed} can be relaxed.}

\begin{figure*}
    \centering
    \includegraphics[width=\textwidth]{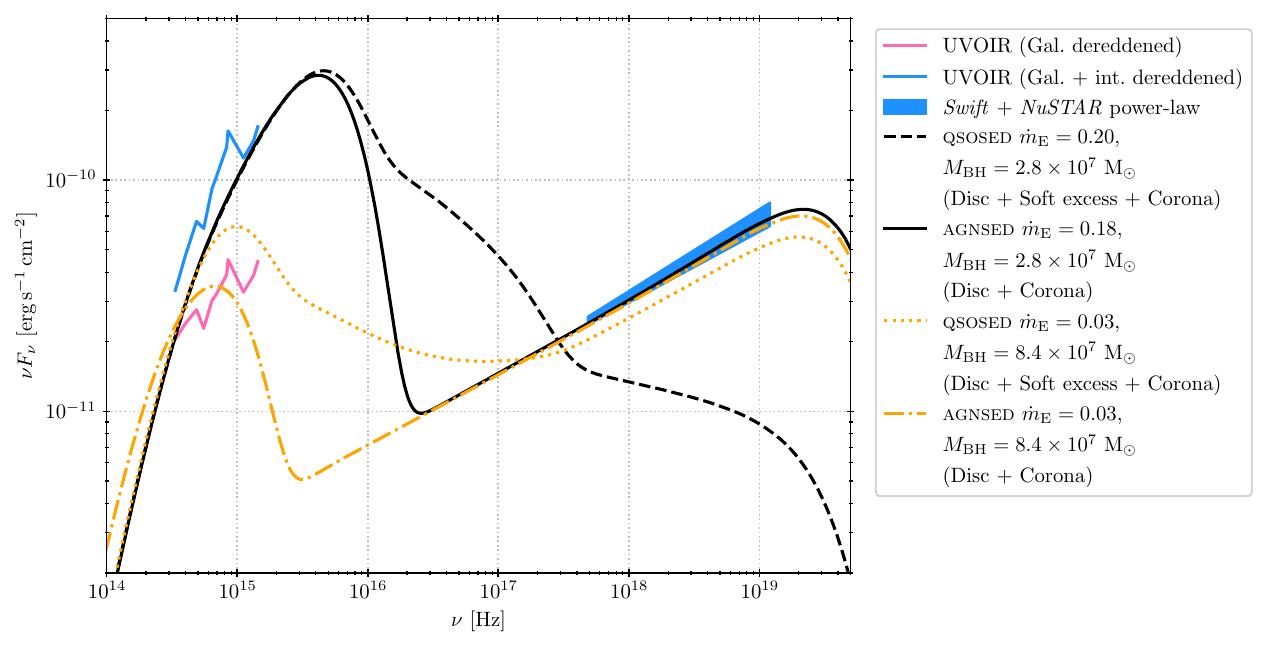}
    \caption{The broadband SED of \mcg.
    Black lines show model SEDs for a fixed black hole mass $2.82\times10^7~\mathrm{M}_\odot$: the dashed black line shows the simple \textsc{qsosed} SED for an accretion ratio $\dot{m}_\mathrm{E}=0.20$;
    the solid black line shows the \textsc{agnsed} SED for $\dot{m}_\mathrm{E}=0.18$, fit to the X-ray spectrum and excluding a soft X-ray excess component (see Section~\ref{sec:bbsed} for details).
    Orange lines show model SEDs for a fixed black hole mass $8.4\times10^7~\mathrm{M}_\odot$: the dotted orange line shows the \textsc{qsosed} SED for $\dot{m}_\mathrm{E}=0.03$;
    the dot-dashed orange line shows the \textsc{agnsed} SED for $\dot{m}_\mathrm{E}=0.03$.
    The pink curve shows the host-galaxy emission subtracted UVOIR spectrum, corrected for Galactic dust reddening;
    the blue curve is additionally corrected for intrinsic reddening as described in Section~\ref{sec:reddening}.
    The blue quadrilateral marks the 3$\sigma$ uncertainty region for a power-law fit to the 2--50~keV X-ray continuum, described in Section~\ref{sec:xspec}.
    }
    \label{fig:full_sed}
\end{figure*}

Fig.~\ref{fig:full_sed} shows the \textsc{qsosed} model for $\dot{m}_\mathrm{E}=0.2$ \textcolor{black}{and $M_\mathrm{BH}=2.82\times10^7~\mathrm{M}_\odot$ (based on the analyses in Sections~\ref{sec:xspec} and \ref{sec:reddening})}.
This model matches the dereddened UVOIR spectrum (blue) but under-predicts the hard X-ray flux considerably, giving $\log(F_{2-10\mathrm{keV}}/\mathrm{erg\,s^{-1}\,cm^{-2}})\approx-10.67$ and has a much softer X-ray spectrum that the observed one.

Using the more flexible model \textsc{agnsed} 
we therefore perform a fit to only the hard X-ray continuum regions \textcolor{black}{as in Section~\ref{sec:xspec} and determine $\Gamma=1.72$ (consistent with the simple power-law model), $\dot{m}_\mathrm{E}=0.18$ and corona radius $R_\mathrm{hot}=22.6$~$R_\mathrm{g}$}. 
In neither our own preliminary X-ray analysis or that of previous researchers (\citealt{Matt06}; \citealt{Bianchi10}; \citealt{Patrick12}) has a soft X-ray excess been found. 
We therefore set the radius of the warm corona equal to that of the hot corona, so that no soft X-ray excess component \textcolor{black}{was} produced.
\textcolor{black}{The resultant model is shown by the solid black line in Fig.~\ref{fig:full_sed} and is a reasonable match to both the X-rays and the dereddened UVOIR spectrum.} 

\textcolor{black}{Although the above \textsc{agnsed} fit does match the X-ray spectral index, the accretion ratio has to be high to account for the X-ray luminosity, and this combination of spectral index and accretion ratio is not seen in the three AGN modelled by \cite{Kubota18} or in the large sample studied by \cite{Hagen24}. It is 
not expected in the original \textsc{qsosed} model where accretion ratio and X-ray spectral index are physically linked. The \textsc{agnsed} fit also requires substantial reddening to reduce the expected UVOIR continuum to the observed levels.} 
We therefore consider next the possibility that the observed UVOIR spectrum (corrected \textit{only} for Galactic reddening – i.e.\ the pink line in Fig.~\ref{fig:full_sed}) is actually the true source spectrum, and that it is not reddened by dust in \mcg.
\textcolor{black}{This possibility requires a low $\dot{m}_\mathrm{E}\approx0.03$. In the \textsc{qsosed} model we also require a higher black hole mass ($8.4\times10^{7}~\mathrm{M}_\odot$) in order to produce the observed X-ray luminosity. This model does, however, produce a weak soft X-ray excess.
A similar, high mass model with \textsc{agnsed}, setting $R_\mathrm{warm}=R_\mathrm{hot}$, is also a reasonable match to the data with the hot inner corona extending to $\approx120~R_\mathrm{g}$, yielding little UV emission and no soft excess.}

\textcolor{black}{In summary, adopting the RM mass $2.82\times10^7$~M$_\odot$ implies that $\dot{m}_\mathrm{E}\approx0.2$ (Appendix~\ref{app:medd})  
and requires substantial intrinsic reddening to explain the observed UVOIR spectrum.
Alternatively, a higher mass ($\approx8\times10^7$~M$_\odot$) and lower $\dot{m}_\mathrm{E}\approx0.03$ can explain the faint, red UVOIR spectrum without additional dust.
This latter scenario may more naturally explain the slope of the hard X-ray spectrum, and the apparent lack of a soft X-ray excess which \citealt{Kubota18} show disappears for $\dot{m}_\mathrm{E}\lesssim0.02$.}

For either mass and accretion ratio scenario, the observed AGN UVOIR spectrum has a shape unlike that expected from a \textcolor{black}{pure} disc, but quite similar to that expected from reprocessing in the BLR \cite[see e.g.][]{Lawther18}. In particular it rises towards longer wavelengths and, although the spectral resolution is limited, it has bumps at 3000--4000~\si\angstrom, and at $\sim7000$--8000~\si\angstrom, where emission from the Balmer and Paschen continua, respectively, are expected. 
Given also the very high correlation between the X-ray and other wavebands, we conclude that the most likely source of the UVOIR variations is the BLR, which has a direct line of sight to the X-ray source. 

\section{Physical Lag Models}
\label{sec:models}
\textcolor{black}{In this Section we test whether the observed lag spectrum can be reproduced by theoretical reprocessing models. 
Using the SED and $\dot{m}_\mathrm{E}$ estimates from Section~\ref{sec:sed}, we evaluate several geometries to determine which physical scenario best explains the wavelength-dependent delays.}
Although the evidence presented above indicates that the most likely source of the UVOIR radiation is the BLR, we also briefly consider here disc reprocessing  (Sec.~\ref{sec:kyn}) which, until recently, was considered to be the `gold standard' model. We also consider reprocessing from a `bowl' geometry (Sec.~\ref{sec:bowl}) which attempts to fit simultaneously both the lag spectra and the SED before concluding   with a short discussion of reprocessing from the BLR in the form of the radiation pressure confined clouds model (Sec.~\ref{sec:blr}).

\subsection{\kynreverb\ disc model}
\label{sec:kyn}
\textcolor{black}{We first examine the standard thin-disc model as a baseline for pure disc reprocessing.}
The fully relativistic UV-optical disc thermal reverberation code package \kynreverb\ has been used by \cite{Kammoun21} to perform an in-depth exploration of AGN accretion disc lag spectra.
\textcolor{black}{To model the \mcg\ lag spectrum we} set most of the parameters to the fiducial values given by \cite{Kammoun21} so that the disc inner radius is determined by the innermost stable circular orbit, the corona height $h=10~R_\mathrm{g}$, and the system inclination is 40$^\circ$.
Following \cite{Kammoun21}, we set the disc colour temperature correction factor to $f_\mathrm{col}=2.4$ across the whole disc.
The colour temperature correction factor modifies both the shape and luminosity of the disc spectrum and was described in detail by \cite{Done12}\footnote{\textcolor{black}{Although the form of $f_\mathrm{col}$ was derived in the context of stellar-mass black hole binary discs, its application was tested against full radiative transfer models for AGN discs and reasonably good agreement between the resultant disc spectra was found: see section 2 of \cite{Done12}.}}.
In turn $f_\mathrm{col}$  modifies the lag spectrum: higher values of $f_\mathrm{col}$ give generally longer lags and greater curvature in the lag spectrum, with a much steeper UV-optical rise (see the discussion in section 4.2 of \citealt{McHardy23}).
As well as fixing $f_\mathrm{col}$ to a specific value across the whole disc, it is also possible to implement a more physically-motivated scheme in which $f_\mathrm{col}$ changes across the disc according to its temperature at each radius (following the prescription of \citealt{Done12}).
After some experimentation, we determine that a constant $f_\mathrm{col}=2.4$ fits our observations best, 
since the additional curvature helps to match both the steep X-ray to UV lag and the flattening of the lag spectrum toward the infrared.
Models without any such correction ($f_\mathrm{col}=1$) or following the \citealt{Done12} prescription were unable to approximate the lags across the full observed wavelength range.

\begin{figure}
    \centering
    \includegraphics[width=0.95\columnwidth]{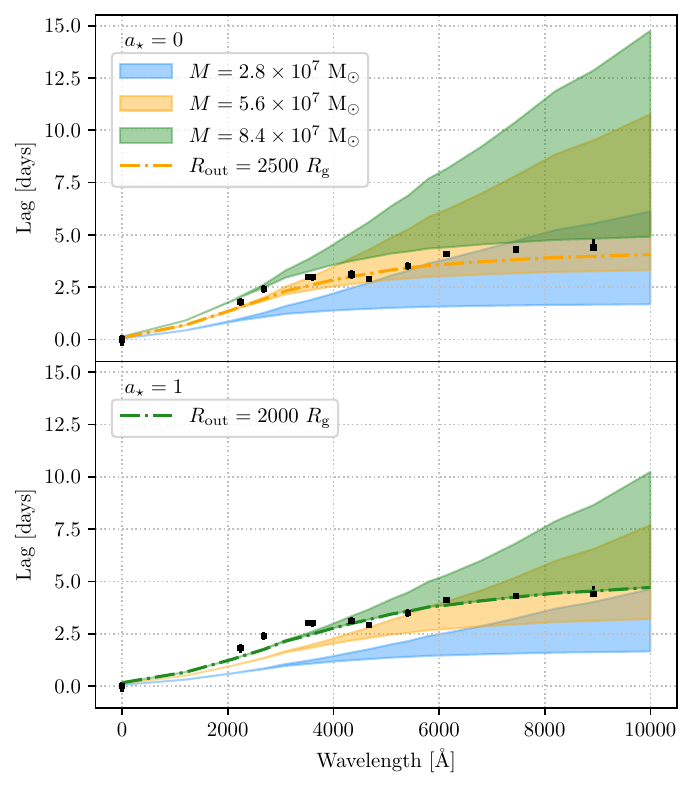}
    \caption{Lag spectra predicted by the model \kynreverb\ (coloured regions) compared with the empirical lags determined by \javelin\ (black points).
    Lag spectra are calculated for the assumed black hole mass $2.82\times10^{7}~\mathrm{M}_\odot$ (blue) as well as masses a factor of 2 and 3 greater (orange and green, respectively).  
    For each mass, lag spectra are generated for both zero spin (top panel) and maximum spin (bottom panel) and for a range of outer disc radii between 2000 and 10000 $R_\mathrm{g}$.
    The dot-dashed orange line shows the predicted spectrum for a mass $5.6\times10^{7}~\mathrm{M}_\odot$ and $R_\mathrm{out}=2500~R_\mathrm{g}$ in the zero spin case and dot-dashed orange line shows the predicted spectrum for a mass $8.4\times10^{7}~\mathrm{M}_\odot$ and $R_\mathrm{out}=2000~R_\mathrm{g}$ in the maximum spin case.
    }
    \label{fig:kynlags}
\end{figure}

We run the models for both spin $a_\star=0$ and 1 and in each case we simulate for a range of outer disc radii between 2000 and 10000~$R_\mathrm{g}$.  The lags for our assumed mass and accretion ratio are shorter than the observed ones. 
The black hole mass is the parameter which has the greatest effect on the scale of the lags, and the estimated mass for \mcg\ is quite uncertain (a factor $\approx3$, \citealt{Fausnaugh17}).
Therefore, as well as performing calculations for the fiducial mass $2.82\times10^{7}~\mathrm{M}_\odot$, we additionally perform runs with the mass increased by a factor 2 and 3.  
The range of predicted lags for each mass and outer disc radius are shown in Fig.~\ref{fig:kynlags}, for both spins $a_\star=0,1$.
For $a_\star=0$, a mass twice as large as the RM estimate and an outer disc radius $R_\mathrm{out}\approx2500~R_\mathrm{g}$ (the orange dashed line in Fig.~\ref{fig:kynlags}) we obtain a reasonable match to the scale and shape of the observed lag spectrum.
Increasing the spin decreases the lag at each wavelength, so for $a_\star=1$ an even higher mass can approximately match the observed lag spectrum with $R_\mathrm{out}\approx2000~R_\mathrm{g}$ (the green dashed line in Fig.~\ref{fig:kynlags}). The fact that higher masses produce lag spectra which are closer to the observations is consistent with the arguments earlier based on the overall SED, the X-ray spectrum, the accretion ratio and the X-ray variability properties which also argue in favour of a higher mass, from a different perspective.

\textcolor{black}{In summary, the RM derived black hole mass appears too low to reproduce the observed lag spectrum using a GR thin disc, lamp-post reprocessing geometry. 
Adopting a higher mass together with a large disc colour-temperature correction can yield lags of comparable overall scale, but the detailed wavelength dependence remains poorly matched. 
In the following subsection, we investigate an alternative disc geometry.}

\subsection{Bowl model}
\label{sec:bowl}
\begin{figure}
    \centering
    \includegraphics[width=\columnwidth]{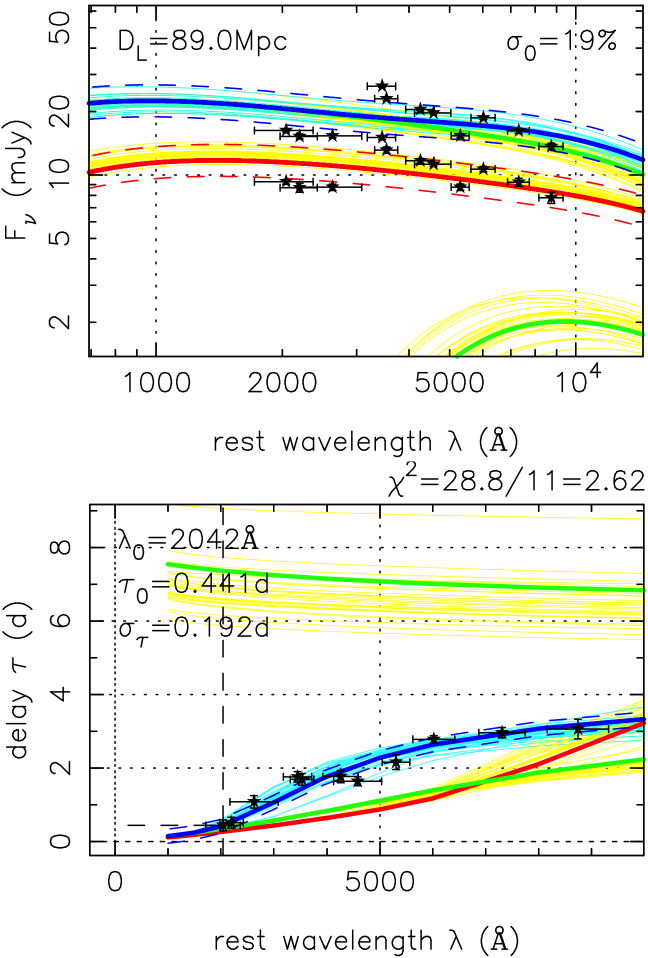}
    \caption{Bowl model analysis.  
    The top panel shows the disc spectrum and the bottom panel shows the model lag spectrum (blue) compared with the measured one (black stars).
    The assumed black hole mass is $2.8\times10^7~\mathrm{M}_\odot$ and the input faint and bright AGN SEDs have been dereddened for both Galactic and intrinsic reddening as described in Section~\ref{sec:reddening}.
    Red and blue curves give SED and lag predictions for the faint and bright states.  
    The green curves give bright-state lags and SEDs separately for the flat disk and for the steep rim, inside and outside of the temperature minimum.
    }
    \label{fig:bowl}
\end{figure}

In the \kynreverb\ model, the combination of a high black hole mass, large colour correction factor and small disc outer radius were required to match the shape of the observed lag spectrum (\textcolor{black}{steep at short wavelengths, flattening at long}).
An\textcolor{black}{other}  way to achieve this shape with a standard disc is with an alternative geometry, such as a `bowl-shaped' accretion flow consisting of a flat disc that flares into a larger scale height wall beyond a certain radius (e.g.\ \citealt{Starkey23}; \citealt{Edelson24}).
We assume the \cite{Fausnaugh17} mass estimate ($2.82\times10^{7}~\mathrm{M}_\odot$) and run the model with the disc inner radius set to $6~R_\mathrm{g}$, the corona height to $10~R_\mathrm{g}$ and the system inclination to $45^\circ$.
The disc rim \textcolor{black}{height} rises with radius as $H \propto R^{100}$.
\textcolor{black}{F}its to our `faint' and `bright' AGN SEDs \textcolor{black}{(}evaluat\textcolor{black}{ed}  at X$_\mathrm{min}$ and X$_\mathrm{max}$ [Section~\ref{sec:fluxflux}] and dereddened as described in Section~\ref{sec:reddening}) are shown in Fig.~\ref{fig:bowl}.
This model predicts \textcolor{black}{a} disc truncat\textcolor{black}{ion} radius of $5.90\pm0.60$~light days ($\approx4000$~$R_\mathrm{g}$) \textcolor{black}{and a temperature there of} $\sim2000$~K, approximately the dust sublimation temperature \citep{Barvainis87}.
The model fit to the faint SED gives an upper limit on the accretion rate $\dot{M}<0.153~\mathrm{M}_\odot$\,yr$^{-1}$ and $L/L_\mathrm{Edd}<0.198$.
The modelled lag spectrum is shown in the lower panel of Fig.~\ref{fig:bowl}.
\textcolor{black}{T}he model is able to reproduce the scale of the observed lags and the overall spectrum shape but still produces a smooth lag-wavelength relationship in contrast to the more complex pattern observed.
A forthcoming version of this model includes an opacity-dependent rim height allowing the Bowl model to produce Balmer and Paschen edges in the lag spectrum (Sinha \& Horne, in preparation).

\textcolor{black}{Since none of the disc models presented so far can account for the \textit{u}- and \textit{i}-band excess lags, 
we next test a model with an additional reprocessing component.}

\subsection{Radiation pressure confined clouds}
\label{sec:blr}
\textcolor{black}{In this subsection we test a model which also includes a reprocessing contribution from the BLR that can introduce additional structure into the lag spectrum.}
\cite{KG01,KG19} have shown that incident high energy emission, reprocessed in the BLR, can give rise not only to strong broad optical--IR recombination lines (the Balmer and Paschen series lines of hydrogen), but also a significant thermal diffuse continuum component (DC), which at UV--optical wavelengths is dominated by hydrogen free-bound recombination continuum (the Balmer and Paschen continuum), with a weaker contribution from free-free emission, and a strong Rayleigh scattering feature  which dominates the diffuse continuum in the vicinity of Ly\,$\upalpha$. 
For example, in Fig.~5 of \cite{KG01}, they show the lags that might be expected in the wavelength range 1000--7000~\si\angstrom\, from NGC\,5548 during two {\it HST} and {\it IUE} monitoring campaigns. 

\begin{figure}
    \centering
    \includegraphics[width=\columnwidth]{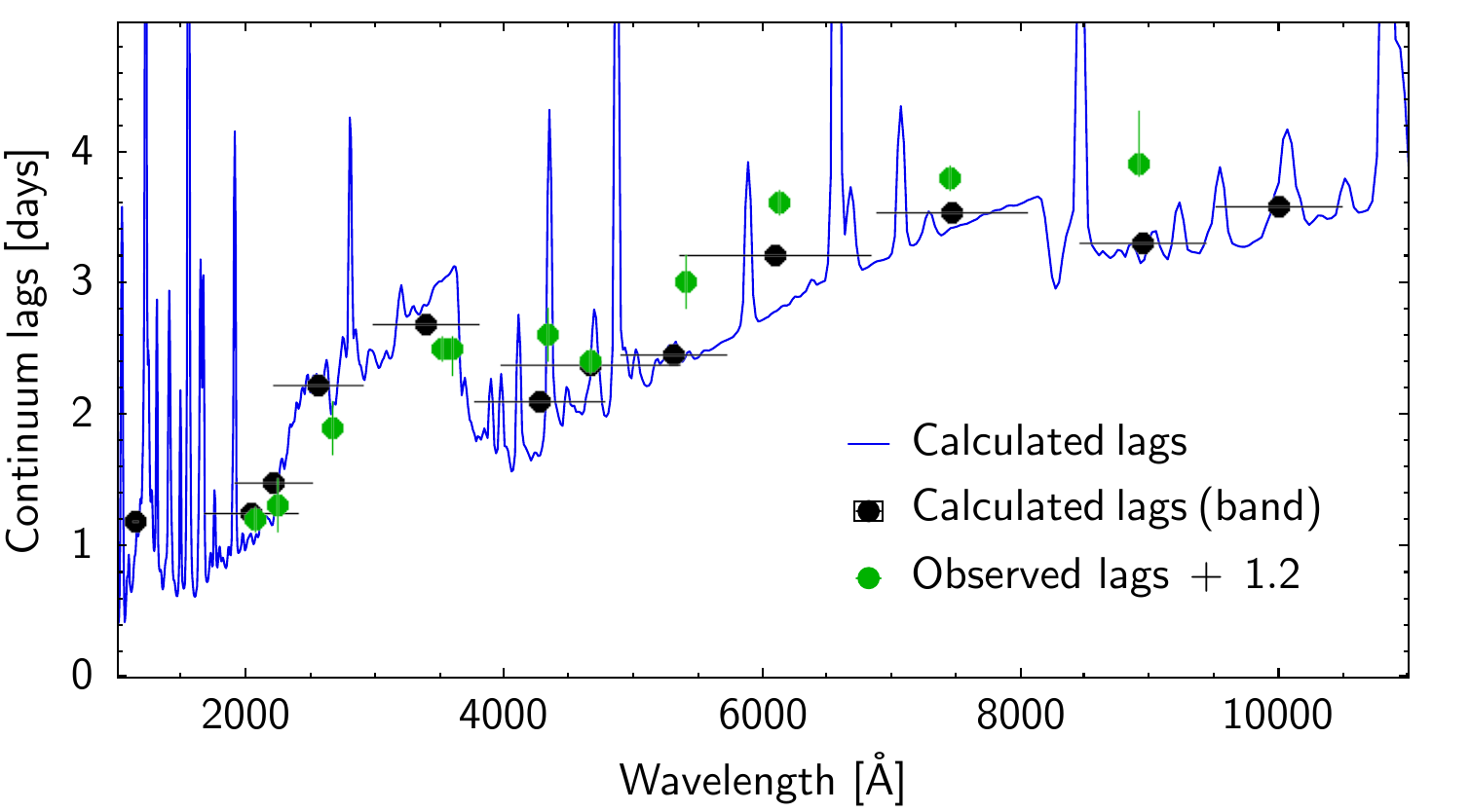}
    \caption{A fit to the lag observed spectrum (green points) with a radiation pressure confined cloud model (similar to the models presented by \protect\citealt{Netzer24} for Mrk\,817).
    The full model lag spectrum (blue) is dominated by the broad line region.
    The mean model lags in several broad bands are shown in black, for easier comparison with the observed ones.}
    \label{fig:rpcc}
\end{figure}

Recently \cite{Netzer20,Netzer22} has calculated similar lags under the assumption that the BLR is made up of radiation-pressure-confined (RPC) clouds. 
The clouds are irradiated by a combination of UV emission from the disc and X-ray emission from the central source. 
\cite{Netzer22} showed that, for a small sample of AGN, most of the observed lags could be reproduced by modelling the response of the BLR clouds, with only a minor contribution from the accretion disc (if any).
A more detailed model of this type, applied to Mrk\,817, is shown and discussed in \cite{Netzer24}.
In Fig.~\ref{fig:rpcc} we show the full model prediction for \mcg\ over all wavelengths as a blue line. Our observed lags are shown as green circles. 
To allow easier comparison to our observations, the model lags averaged over the wavelength ranges of our observational bands are shown in black.
In this RPC cloud model, the lags are calculated relative to the ionising UV radiation and assuming an offset of 1~day relative to the 1100~\si\angstrom\ continuum. 
In Fig.~\ref{fig:rpcc} the observed lags have been shifted by 1.2~days so that the shortest-wavelength lag (UVW2 2084~\si\angstrom) approximately matches the relevant calculated lag in that band. 

Over the UVOIR range covered by the modelling, and normalised to the UVW2 lags, the RPC cloud model and observed lags are in reasonable agreement. 
They agree particularly well in the regions of the Balmer and Paschen continua.
 
\textcolor{black}{Taken together, the modelling in this Section shows that a thin disc alone cannot reproduce the observed lag spectrum.
The excess lags around \textit{u} and \textit{i} are highly suggestive of reprocessing by gas in the BLR.
The flattening of the lag spectrum towards longer wavelengths may be due to a high colour temperature correction factor, or the onset of a large scale-height disc wind or flared disc.
We discuss the implications our findings below in Section~\ref{sec:discussion} and conclude in Section~\ref{sec:conclusions}.}

\section{Discussion}
\label{sec:discussion}
In this paper we have presented 12 very high-quality lightcurves of the bright Seyfert 1.5 AGN \mcg\ from X-ray to near-IR.
Over a three month period we obtained on average more than two observations per day with \swift\ (X-ray and UV/optical) and better than daily cadence with ground-based observatories (optical and near-IR bands). From these data we have determined one of the best-defined AGN lag spectra yet produced.
We measured lags relative to the \swift\ UVW2 band using two different methods (ICCF and \javelin) and obtained very consistent results. 
From flux-flux analysis of the lightcurves we have distinguished the separate SEDs of a constant galaxy component and a variable AGN component. We have analysed the combined \swift-XRT data from our observations with archival \nustar\ data to provide a 0.3--50 keV spectrum which we have combined with the UVOIR AGN data to provide a multiband SED. 

Many of the observational results are common to those from similar campaigns on other AGN.
These include:
(i) UVOIR lags which increase with wavelength;
(ii) a lag spectrum that is \textcolor{black}{flatter} than predicted by a simple disc reprocessing model; 
(iii) an X-ray/UV lag that is longer than one would expect from an extrapolation of the UVOIR lags (see Fig.~\ref{fig:lag_spec}); and
(iv) `excess' lags in bands covering the Balmer and Paschen jumps.
However, in strong contrast to many previous studies, the X-ray and UVOIR lightcurves of \mcg\ are highly correlated.
We find $r_\mathrm{max}=0.85$ between the X-ray and UVW2 bands which is, to the best of our knowledge, the strongest such correlation yet reported for an AGN.
Additionally, whilst previous, similar studies have found that the variable component has a shape consistent with a canonical, blue disc spectrum (e.g.\ \citealt{Cackett20}; \citealt{McHardy23}), here we find a much redder variable spectrum. 
If the variable spectrum relates to the AGN accretion disc, then this finding implies that either the disc is cool (likely requiring a high black hole mass), or that the emission is substantially reddened by dust within the host galaxy.
Below, we briefly discuss the most important findings from this study.

\subsection{The shape of the lag spectrum}
\textcolor{black}{In Section~\ref{sec:timing} we determined the lags between photometic bands using two methods: ICCF and \javelin. The lags produced by the two methods are almost identical. Particularly for our very well sampled data, we find no reason why the assumptions built into \javelin, such as interpolation assuming DRW variability, should produce incorrect lags. However we discuss this matter in more detail in Appendix~\ref{app:javelin}}.

We \textcolor{black}{then} modelled the lag spectrum with disc and bowl models. Both require an edge, e.g.\ produced by an obscuring gaseous or perhaps dusty disc wind, to flatten the lag spectrum at longer wavelengths. The disc model requires a high colour correction factor of 2.4 to introduce sufficient curvature in the lag spectrum  so that it can fit both the rapid increase in lags at short wavelengths and the slower increase at long wavelengths. The effect of a disc edge and of the colour correction factor are discussed in detail following observations of NGC\,4395 by \cite{McHardy23}. We also refer readers to \cite{Kammoun21} where the effects on the lag spectrum of the variation of many parameters, including the outer disc radius, are discussed. We find that, to obtain a good fit to the lags with the disc model, a mass of 2 or 3 times greater than the optical RM mass of \cite{Fausnaugh17} of $2.82 \times 10^{7}$\msun, is required (i.e.\ at the upper end of the RM mass uncertainty range). 

Both disc and bowl geometries can provide a reasonable description of the overall scale and shape of the lag spectrum but they cannot explain the most prominent features, i.e.\ the excess lags in the $u$ and $r-i$ bands. These excesses are, however, easily explained as the Balmer and Paschen continua, arising from reprocessing in the BLR gas \citep{KG01,KG19,Lawther18,Netzer20,Netzer22}. Here we obtain a good fit using the radiation-pressure-confined cloud model of \cite{Netzer22}. Reprocessing in the BLR, which is more distant than the disc regions associated with UVOIR reprocessing, has the advantage that the large lag between the X-ray variations and the UV variations, which is hard to explain in disc reprocessing without the maximum 2.4 colour correction factor, is automatically explained. 

\subsection{The shape of the broadband SED}
In Section~\ref{sec:xspec} we modelled the time-averaged X-ray spectrum using data from our campaign, together with archival \nustar\ spectra observed at a similar time. We measured the X-ray spectral index up to 50~keV, finding a hard spectrum with $\Gamma=1.697\pm 0.008$ and determined the 2--10~keV X-ray luminosity.
In Section~\ref{sec:fluxflux} we used the flux-flux method to decompose the multiband photometry into constant and variable components, taken to represent the host galaxy and AGN emission, respectively.
The lack of curvature in the flux-flux plots is most easily explained if the spectral shape of the variable component does not vary with luminosity. Similar
linear flux-flux plots have been shown in many other studies (e.g.\ \citealt{Fausnaugh16}; \citealt{Starkey17}; \citealt{McHardy18}; 
\citealt{HS20}; 
\citealt{Vincentelli21}; 
\citealt{Prince25}).
However \cite{Netzer24} showed that variations in \textit{HST} STIS spectra in  Mrk\,817
were consistent with a bluer-when-brighter interpretation of the variable component, although \cite{Cackett23} had previously found no curvature in the flux-flux plots derived from \swift\ lightcurves. In the case of \mcg\, there is no requirement for spectral variability in the variable component to explain our flux-flux plots. Therefore, as we do not have any spectra to allow for deeper investigation, we proceeded with the flux-flux decomposition as the best available bespoke determination of its AGN and host galaxy spectra.

In Section~\ref{sec:reddening} we noted that the AGN UVOIR spectrum obtained from our flux-flux analysis was much fainter and redder than the canonical disc spectrum of a bright AGN of this RM mass.
The soft X-ray spectrum of \mcg\ is known to be heavily absorbed beyond purely Galactic extinction and absorption features are seen in optical resonance lines, 
both indicating the presence of subtantial gas within the host galaxy along our line of sight.
Since dust may also be present along with this gas, we attempted to `correct' the UVOIR spectrum for reddening due to dust inside \mcg.
After some experimentation, we found that extinction correction using a flat reddening curve with intrinsic $A_V=1.05$~mag could correct the flux-flux UVOIR spectrum to a blue disc spectrum with $\dot{m}_\mathrm{E}\approx0.2$.
(We note that the determination of the intrinsic UVOIR SED is quite uncertain, since it incorporates both the limitations of the flux-flux analysis to produce the `observed' spectrum, discussed above, and uncertainty in the dereddening procedure mentioned in Section~\ref{sec:reddening}.)

In Section~\ref{sec:bbsed} we combined the UVOIR and X-ray spectra to model the broadband SED.
We used the simple, prescriptive \textsc{qsosed} model  to determine the expected SED for the RM mass and $\dot{m}_\mathrm{E}\approx0.2$.
By design this SED matches the Galactic $+$ intrinsic dereddened UVOIR spectrum well, but it completely fails to match the X-ray data, predicting a much fainter and softer X-ray spectrum.
It is possible to obtain a good fit to the broadband data using the more flexible model \textsc{agnsed}, in which the X-ray photon index is a free parameter.
This model is shown by the solid black line in Fig.~\ref{fig:full_sed}.
This solution is energetically possible as the variable UVOIR power is less than the variable driving X-ray power, although the SED shape is unusual.  

It has been established that the X-ray photon index is positively correlated with \me\ \citep[e.g.][]{Risaliti09,Kubota18}\footnote{However, see \cite{Chen25} for a discussion of the significance of this correlation.} and this relationship is hard-wired into the \textsc{qsosed} model.
The hard X-ray photon index we measure, $\Gamma=1.697$, suggests a low accretion ratio of $\dot{m}_\mathrm{E} \approx 0.03$ \citep{Risaliti09}, similar to the estimate of 0.05 from \cite{Fausnaugh16} based on the 5100~\si\angstrom\ luminosity. 
Given the inferred bolometric luminosity of \mcg\ (from hard X-rays and other indicators: see Appendix~\ref{app:medd}), this much lower value of \me\ is only possible if the black hole mass is in fact a factor of a few greater than the RM estimate of \cite{Fausnaugh17}.
A higher mass and low \me\ naturally explain the \textit{observed} SED shape (without requiring substantial additional reddening) since the intrinsic UVOIR spectrum will be fainter and redder as the result of a dimmer and colder disc, and the expected X-ray spectrum will be much harder.

Given the factor $\approx3$ uncertainty in the RM black hole mass, we produce SED models for a mass at the upper end of this range, and $\dot{m}_\mathrm{E} \approx 0.03$.
The \textsc{qsosed} prediction (the dotted orange line in Fig.~\ref{fig:full_sed}) does not quite match the data (the UVOIR spectrum corrected only for Galactic reddening, shown in pink): being too bright and blue in the UVOIR band and slightly underluminous in X-rays.
With \textsc{agnsed} we are able to increase the radius of the hot, inner accretion flow to increase the X-ray luminosity at the expense of the UV, thereby achieving a better match to the data.
This best overall fit, constrained mainly by the X-ray part of the SED, adopts a mass of $8.4 \times 10^{7}$\msun\, a factor $\approx3$ greater than that of \cite{Fausnaugh17}, but consistent with some other indicators such as the X-ray variability properties (see Section~\ref{sec:xvar}). 
\cite{Kubota18} find that an accretion ratio of \me$\sim0.02$, close to our current best estimate, is the approximate limit below which the soft excess and strong blue disc emission disappear, which may explain our observations\footnote{In addition the high line-of-sight Galactic column density, $N_\mathrm{H}\approx2\times10^{21}$~cm$^{-2}$, is a hindrance to the detection of the soft excess.} \citep[e.g.][]{Matt06,Patrick12}. The lack of a strong disc component is perfectly consistent with the deduction, from the shapes of both the lag spectrum and the AGN SED as deduced by flux-flux analysis, that the large part of the variable UVOIR emission is reprocessed radiation from the BLR.

So, in summary, our analysis for Fig.~\ref{fig:full_sed} presents two possibilities - either: 
\vspace*{-2mm}
\begin{enumerate}
    \item the RM mass is accurate, so the bolometric luminosity implies $\dot{m}_\mathrm{E}\approx0.2$,
    for which the observed X-ray spectrum would be unusually hard and bright. Luminous UVOIR disc emission would be expected and so, to reduce that luminosity to the observed levels, 
    substantial intrinsic reddening (above Galactic) would be required. Or,
    \item the X-ray photon index is a reliable indication of a low accretion ratio ($\dot{m}_\mathrm{E}\approx0.03$) which requires that the black hole mass is at the upper end of the uncertainty range on the RM estimate. The broadband SED is closer to the expectations for an AGN of this (higher) mass and (lower) \me.  The expected UVOIR luminosity would be much lower, consistent with the observed values without any reddening above Galactic and consistent with a large fraction of the UVOIR luminosity arising from reprocessing of X-rays in the BLR. 
\end{enumerate}

\subsection{Proposed geometry}
We propose that \mcg\, is not, as was thought previously, a high accretion ratio AGN but is a low accretion ratio AGN ($\dot{m}_\mathrm{E}\approx0.03$) with a relatively high mass, approaching $10^{8}$~\msun.
We found that the AGN UVOIR spectrum was faint and red, whilst the X-ray emission was relatively luminous.
Without invoking reddening within \mcg\ to explain the faint, red UVOIR spectrum we can instead approximate the broadband SED with a model in which the standard (geometrically thin, optically thick) disc does not extend all the way down to the innermost stable circular orbit.
Instead, the inner flow is taken up by a relatively large, hot, X-ray emitting region, meaning the standard disc is intrinsically UV-faint. 
The very high correlation between the X-rays and other wavebands shows that there is a direct line of sight to the reprocessor(s) and the shape of the lag spectrum and AGN SED show that the main reprocessor is the BLR. The hard X-ray spectrum is strongly absorbed below $\approx2$~keV. 
Additionally, brief, sharp drops in X-ray flux are observed along with a hardening of the 0.3--10keV spectrum (e.g.\ Fig.~\ref{fig:fian}), likely as the result of increased obscuration by dense gas on small spatial scales.
At least part of the absorption then probably occurs near the inner radius of the BLR where the gas density is high.
The absorbing gas may be part of a failed radiatively accelerated dusty outflow \citep{Czerny11}: around the dust sublimation radius in the disc, a dusty wind is launched. The transition to dust-free clouds marks the onset of the inner BLR. Alternatively the wind could be a line-driven gaseous wind \citep{Proga04} as suggested for NGC~4395 \citep{McHardy23}.
The large scale height of the outflow/BLR presents a large solid angle to the primary continuum source, leading to a high luminosity in reprocessed radiation, and strong correlations between the driving and reprocessed continuum lightcurves.

This geometry is different to that suggested for NGC\,5548 and NGC\,4151 despite these two AGN having similar masses and accretion ratios to \mcg\ ($M_\mathrm{BH}\sim4\times10^{7}$~\msun\ and $\dot{m}_\mathrm{E}\sim0.01$: \citealt{Edelson17}, \citealt{Kubota18}).
Intensive \swift\ monitoring of other AGN, including these, has revealed that the delay of UV variations behind X-rays is often longer than predicted by a simple extrapolation of the UVOIR lag spectrum to the X-ray band (e.g.\ \citealt{Edelson17,McHardy18,Edelson19}).
For both NGC\,5548 and NGC\,4151 an inflated inner accretion disc has been invoked which might scatter and delays the incident X-rays, reprocessing them into far-UV, which then illuminates the outer `standard' disc
(\citealt{GD17}; \citealt{Edelson17}).
Similarly, in \mcg\ we see an X-ray/UV lag ($\approx2$~days) that is too long to be plausibly interpreted as simply the light travel time between the corona and inner accretion disc. However, an intermediate reprocessor results in weak-to-modest X-ray/UV correlations that \textit{are} seen in NGC\,5548 and NGC\,4151 ($r_\mathrm{max}\approx0.4$--0.7) but \textit{not} in \mcg\ in which the X-ray/UV correlation is very strong ($r_\mathrm{max}\approx0.85$), requiring direct illumination of the reprocessor (BLR) by the X-rays.

We find evidence for multiple reprocessing components in \mcg, with different components dominating at different times, possibly linked to changes in luminosity. \cite{Fian23} monitored \mcg\, for six months. Their 2019–20 lightcurves show large-amplitude, long-timescale variations (Fig.~\ref{fig:fian}), unlike ours (Fig.~\ref{fig:lcs}).
\citeauthor{Fian23} also reported longer optical lags than those we measured and a luminosity four times higher. Detrending lightcurves can remove biases in the measured lags \citep{Welsh99} and reveal short-term lags that would otherwise be obscured (e.g.\ \citealt{McHardy14}; \citealt{McHardy18}; \citealt{Pahari20}; \citealt{Vincentelli21}; \citealt{Lewin23}; \citealt{Edelson24}). Detrending the lightcurves of \cite{Fian23} did indeed reveal shorter lags which are consistent with ours. 
Detrending was unnecessary for our lightcurves, as they showed no significant long-term trends.
This analysis shows that both short and long lags were present \textit{concurrently} during the 2019–20 campaign although with the longer lags dominating during that brighter phase. However the shorter lags dominated during our fainter 2021 observations.  A fourfold increase in luminosity could double the inner radius of the BLR (e.g.\ \citealt{Bentz13}), leading to longer lags and smoother lightcurve variations and, if due to an increase in accretion rate, the inner disc radius might also reduce, allowing for shorter disc lags. Similar luminosity-dependent lag behavior has been observed in other AGN, such as in Mrk\,110 \citep{Vincentelli21,Vincentelli22}, where two variable components (attributed to the disc and BLR) showed increasing lags with higher ionising luminosity, consistent with an expanding BLR.

\section{Conclusions}\label{sec:conclusions}
We monitored the AGN \mcg\ intensively for a period of 3 months with \swift\ (in X-rays and UV/optical bands) and ground-based observatories (in optical and infrared bands).
The correlations between all observed wavelengths were strong and in particular the X-ray/UV correlation ($r_\mathrm{max}=0.85$) is the strongest yet reported for an AGN.
This implies an unimpeded line of sight between the X-ray corona and the UVOIR reprocessor(s).

Throughout our monitoring period the target was relatively faint and the UVOIR lags we measured were shorter than those measured in a recent optical campaign during which the source was $\approx4$ times brighter \citep{Fian23}.
We found that short lags (consistent with those we measured) were also present in those earlier data, after filtering out the longer-term trends.
These observations demonstrate that more than one source of UV/optical variability is present within \mcg: at least one of these sources must be associated with reprocessing in the BLR, giving rise to the structure seen on short timescales in both the lag spectrum and the rms flux spectrum. The origin of the longer term variations is unknown but may be associated with the temperature fluctuations observed in AGN accretion discs by \cite{Neustadt22}.

We found that pure disc reprocessing models could not reproduce the complex shape of the lag spectrum, whereas a radiation pressure confined cloud model (approximating the reprocessed emission from the BLR) broadly matched the observations.
If indeed emission from the BLR dominates the lags, the long ($\approx2$~day) lag between X-ray and UV lightcurves is more easily explained than in the disc reprocessing scenario which would require either an extreme corona height or an intermediate reprocessor, which is inconsistent with the very strong X-ray/UV correlation. 

If the optical RM black hole mass estimate of \mcg\ ($M_\mathrm{BH}=2.82\times10^7$~\msun) is accurate, then its bolometric luminosity implies an Eddington accretion ratio $\dot{m}_\mathrm{E}\approx0.2$ which should produce a high UV disc luminosity which is not seen. In this case, much of the UV disc emission must be `hidden' by substantial reddening within \mcg. The X-ray spectrum should also be much softer and less luminous than is observed \citep{Kubota18}. 
Alternatively, if the hard X-ray spectrum is a reliable indicator of the accretion ratio, it suggests a much lower $\dot{m}_\mathrm{E}\approx0.03$ in which case the black hole mass must be at the upper end of the RM mass uncertainty range (i.e.\ $M_\mathrm{BH}\gtrsim8\times10^7$~\msun). This low accretion ratio is consistent with the value (\me$\sim0.02$) below which \cite{Kubota18} predict that the inner optically thick part of the disc disappears, drastically reducing the UVOIR contribution from any remaining parts of the disc. This prediction is in good agreement with our conclusion that the large majority of the shorter timescale UVOIR variability comes from reprocessing in the BLR.

\section*{Acknowledgements}
Thanks go to Carina Fian for providing the 2019-20 Wise optical lightcurves.
DK and I{\Mc}H acknowledge support from the UK Science and Technology Facilities Council (STFC) from grant ST/V001000/1.
EMC and JAM gratefully acknowledge for analysis of the Zowada Observatory data from the NSF through grant AST-1909199. F.M.V. ackowledges finiancial support from the European Union’s Horizon Europe research and innovation programme through the Marie Sk\l{}odowska-Curie grant agreement No. 101149685. 
This work made use of data supplied by the UK Swift Science Data Centre at the University of Leicester.
This work makes use of observations from the Las Cumbres Observatory global telescope network.

\section*{Data Availability}
Data on which this work is based are available from the \swift\ archive\footnote{\url{https://www.swift.ac.uk/swift_live/index.php}} and the LCO archive\footnote{\url{https://lco.global/documentation/data/archive/}}.



\bibliographystyle{mnras}
\bibliography{refs} 



\appendix
\section{Flux-flux analysis procedure}\label{sec:fluxflux-details}
In Section~\ref{sec:fluxflux} we separated the variable and constant spectral components by performing a flux-flux analysis.
We initially calculate the mean spectrum $A_\nu(\lambda)$ and the rms spectrum $S_\nu(\lambda)$ from the data and its uncertainties $F_\nu(\lambda,t)$ and $\sigma(\lambda,t)$, respectively.
We define 
\begin{equation}
    \mathrm{X}(\lambda,t) \equiv \frac{ F_\nu( \lambda, t ) - A_\nu( \lambda ) } { S_\nu( \lambda ) }
\end{equation}
and estimate the lightcurve shape $\mathrm{X}(t)$ by taking a weighted average over all wavelengths:
\begin{equation}
    \mathrm{X}(t) = \sum \mathrm{X}(\lambda,t)W(\lambda,t) / \sum W(\lambda,t),
\end{equation}
where the inverse-variance weights $W(\lambda,t)=(S_\nu[\lambda]/\sigma[\lambda,t])^2$.
$\mathrm{X}(t)$ is then scaled such that $\langle \mathrm{X} \rangle = 0$ and $\langle \mathrm{X}^2 \rangle = 1$.
Next, $\mathrm{X}(t)$ is fixed so $A_\nu(\lambda)$ and $S_\nu(\lambda)$ can be estimated by linear regression.
The algorithm then iterates through calculating $\mathrm{X}(t)$ assuming $A_\nu(\lambda)$ and $S_\nu(\lambda)$ are known, then calculating $A_\nu(\lambda)$ and $S_\nu(\lambda)$ assuming $\mathrm{X}(t)$ is known, until convergence. 

We can then extrapolate the flux-flux relations to a lower brightness level than sampled by the data.
The vertical line $\mathrm{X}_\mathrm{gal}$ on Fig.~\ref{fig:fluxflux} indicates the point at which the flux in the bluest band (UVW2) is just 1$\sigma$ above zero.  Reading off the other band fluxes at $\mathrm{X}_\mathrm{gal}$, we are able to obtain an estimate of the host galaxy SED (shown in dark red in the right-hand panel of Fig.~\ref{fig:fluxflux}).
The average AGN SED is then the average component minus the host galaxy SED (this is shown in pink in the right-hand panel of Fig.~\ref{fig:fluxflux}). 
The resulting SEDs for \mcg\ are given in Table~\ref{tab:fluxflux}.

\begin{table*}
\centering
\caption{Results from the flux-flux analysis in Section~\ref{sec:fluxflux}}
\begin{tabular}{cc|cccccccc}
\hline
Filter & $\lambda_\mathrm{eff}$ & $A_\nu$ & $S_\nu$ & min.\ & max.\ & AGN & Galaxy & Dered$_\mathrm{Gal}$ & Dered$_\mathrm{int}$ \\
(1)    & (2)                    & (3)     & (4)     & (5)   & (6)   & (7) & (8)    & (9)                  & (10) \\
\hline
UVW2 & 2083.95 & $0.622\pm0.002$ & $0.088\pm0.002$ & $0.491$ & $0.849$ & $0.608\pm0.012$ & $<0.012$ & 5.083 & 3.833 \\
UVM2 & 2245.03 & $0.685\pm0.003$ & $0.085\pm0.003$ & $0.559$ & $0.904$ & $0.586\pm0.021$ & $0.099\pm0.021$ & 4.952 & 3.822 \\
UVW1 & 2681.67 & $1.483\pm0.004$ & $0.145\pm0.004$ & $1.269$ & $1.855$ & $0.994\pm0.026$ & $0.490\pm0.026$ & 2.958 & 3.798 \\
U    & 3520.88 & $3.288\pm0.005$ & $0.334\pm0.005$ & $2.793$ & $4.149$ & $2.298\pm0.037$ & $0.990\pm0.037$ & 2.318 & 3.606 \\
$u$  & 3608.04 & $2.902\pm0.007$ & $0.298\pm0.008$ & $2.461$ & $3.667$ & $2.045\pm0.055$ & $0.856\pm0.055$ & 2.294 & 3.569 \\
B    & 4345.28 & $3.931\pm0.006$ & $0.331\pm0.006$ & $3.442$ & $4.782$ & $2.272\pm0.043$ & $1.659\pm0.042$ & 2.042 & 3.211 \\
$g$  & 4671.78 & $5.129\pm0.005$ & $0.357\pm0.005$ & $4.601$ & $6.047$ & $2.452\pm0.036$ & $2.677\pm0.036$ & 1.920 & 3.047  \\
V    & 5411.45 & $6.504\pm0.009$ & $0.350\pm0.009$ & $5.986$ & $7.404$ & $2.404\pm0.066$ & $4.101\pm0.065$ & 1.720 & 2.701 \\
$r$  & 6141.12 & $11.483\pm0.009$ & $0.512\pm0.011$ & $10.724$ & $12.801$ & $3.521\pm0.079$ & $7.962\pm0.079$ & 1.602 & 2.410  \\
$i$  & 7457.89 & $11.694\pm0.016$ & $0.605\pm0.020$ & $10.797$ & $13.251$ & $4.159\pm0.137$ & $7.535\pm0.136$ & 1.437 & 1.972 \\
$z$  & 8922.78 & $14.230\pm0.027$ & $0.686\pm0.031$ & $13.215$ & $15.994$ & $4.711\pm0.217$ & $9.519\pm0.215$ & 1.297 & 1.627 \\
 \hline
\end{tabular}
    \parbox{16cm}{All wavelengths (in \si\angstrom) and fluxes (in mJy) are given in the observed frame.
    Column (3) is $A_\nu$, the mean spectrum; Column (4) is $S_\nu$, the rms spectrum; Columns (5) and (6) are the minimum and maximum total fluxes; Column (7) is the AGN component and Column (8) is the constant host galaxy component. 
    Column (9) gives the multiplicative factor to be applied to the observed fluxes to correct for Galactic reddening ($A_V^\mathrm{MW}=0.568$~mag using the curve of \protect\citealt{CCM89}).
    Column (10) gives the multiplicative factor to be applied to the observed fluxes to correct for intrinsic reddening in the host galaxy rest frame ($A_V^\mathrm{int}=1.05$~mag using the curve of \protect\citealt{Gaskell04}: see Section~\ref{sec:reddening} for details).}
\label{tab:fluxflux}
\end{table*}

\section{ICCF and \javelin}
\label{app:javelin}
\textcolor{black}{In Section~\ref{sec:timing} we determined the lags between photometric bands using two methods: ICCF and \javelin. Here we briefly comment on the comparison between those methods.}

\textcolor{black}{To interpolate between gaps in the observations, \javelin\, assumes that the variations in the light curves can be described by a Damped Random Walk (DRW) model. The power spectral density (PSD) of a variability process that is described by a DRW is flat at low frequencies, bending to a power law slope of index $-2$ at high frequencies. The DRW high frequency slope is consistent with AGN observations in both the X-ray \citep[e.g.][]{McHardy04,McHardy06} and Section~\ref{sec:xvar} here, and optical \citealt{Mushotzky11}; \citealt{Guo17}) bands. However AGN X-ray PSDs have power law slopes of about $-1$ at low frequencies \citep{McHardy04,McHardy06} and we are starting to see the same low frequency slopes in optical PSDs \citep{Beard25}. However our present data are very well sampled and any interpolation occurs within the high frequency slope range of the PSDs so the use of a DRW should introduce no errors.} 

\textcolor{black}{\javelin\, also assumes that the delayed lightcurves are smoothed (by a top-hat function), linearly scaled and shifted versions of the driving lightcurve. The top-hat function would be the correct function if the X-ray source were surrounded by a spherical reprocessor. Whilst that will not be the case for a pure disc reprocessor, the BLR as a reprocessor would be close to that geometry. Therefore, in the present case, any uncertainty introduced by uncertainty in the reprocessing function is likely to be small. As noted above, the ICCF, which merely maximizes the correlation between two lightcurves and makes no assumptions about the form of the variability, gives results which are completely consistent with those of \javelin\, (see Fig.~\ref{fig:lag_spec}). We therefore assume that the assumptions which are built into \javelin\, do not substantially affect our estimated time lags.}

\section{Detrending of the 2019-20 Wise Observatory lightcurves}\label{app:detrend}
\textcolor{black}{
In Section~\ref{sec:fian} we reanalysed the 2019-20 Wise Observatory lightcurves, taken a year prior to our \swift\ campaign. 
To investigate whether the longer lags seen in the un-detrended Wise data are due to an additional longer timescale Fourier component of the lag or whether they are due to an independent cause, we have measured how the lags change depending on the timescale used in the detrending. 
For this investigation we used simple moving-average detrending. 
We first smoothed the input lightcurves with a simple boxcar and then subtracted the smoothed lightcurve from the original. 
We performed this investigation between the 4250~\si\angstrom\ and 7320~\si\angstrom\ lightcurves. 
We chose the 7320~\si\angstrom\ rather than 8025~\si\angstrom\ lightcurve as it is of higher S/N.}

\begin{figure}
    \centering
    \includegraphics[width=8cm,angle=0]{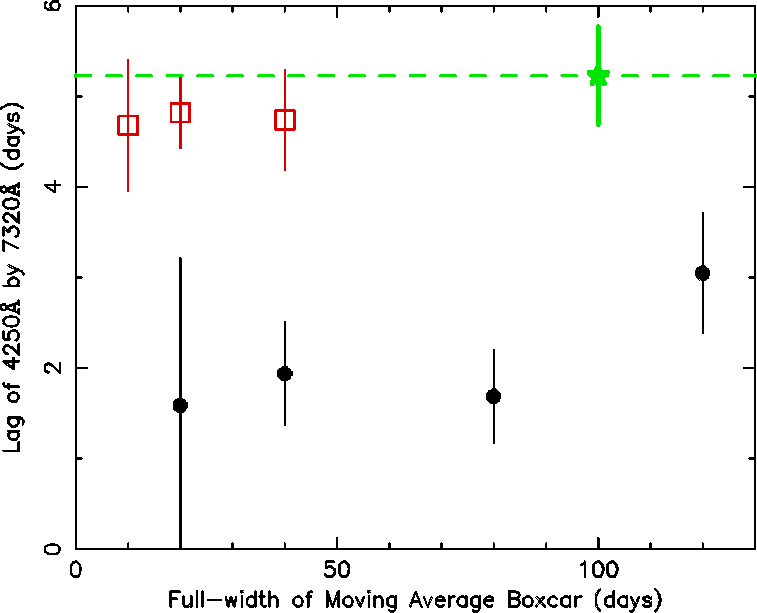}
   \vspace*{-2mm}
   \caption{Lags resulting from detrending the 4250~\si\angstrom\ and 7320~\si\angstrom\ Wise lightcurves with a moving-average boxcar. 
   The filled black circles are the lags from the detrended lightcurves and the open red squares are the lags from the smoothed lightcurves. 
   The dashed green line is the lag measured from the original un-detrended lightcurves with the uncertainty shown by the single green errorbar, whose position on the \textit{x}-axis has no relevance.
   }
    \label{fig:movavg}
\end{figure}

\textcolor{black}{In Fig.~\ref{fig:movavg} we show, as filled black circles, the lags from the detrended lightcurves as a function of the full-width of the boxcar. 
We have also measured the lags between the smoothed lightcurves, shown as open red squares. 
We do not include lags from lightcurves smoothed with boxcars larger than 40~days as there would be less than 5 independent data points in the lightcurves. 
We can, however, use the subtracted lightcurves resulting from longer timescale moving average subtraction as the number of independent data points does not change. 
We also indicate, as a dashed green line, the lag measured from the original, un-detrended, lightcurves. 
The uncertainty on that lag is shown by the errorbar on a single green star. The abscissa value of that star has no meaning. 
We note that, for boxcars up to 80~days, the subtracted lags remain at about 1.5~days, exactly the same value as we obtained from the Lowess filtering with 45~days filtering time. 
At the longest, 120~days, boxcar, the lag rises as the effect of the boxcar becomes minimal. Meanwhile, the lag from the smoothed lightcurves are all longer, around 4.8~days, just below the lag measured from unfiltered lightcurves.}

\textcolor{black}{If we can assume that the short lags (1.5~days) measured in the detrended lightcurves are due to reprocessing, then the fact that these lags do not change even when timescales as long as 80~days remain in the detrended lightcurves indicates that the short lags are not the result of removing longer timescale Fourier components from the lag reprocessing function. 
The longer lags which we see in the undetrended lightcurves and in the smoothed lightcurves, are therefore probably the result of an independent process. 
We do not know what that process might be but \cite{Neustadt22} show that temperature fluctuations can move both inward and outward across the accretion disc producing both positive and negative lags on longer timescales than reprocessing lags ($\sim100$~days, similar to those observed for \mcg: \citealt{Fian23}; Fig.~\ref{fig:fian}). 
Moreover, the resultant slow flux variations dominate in amplitude over reprocessing variations. This may be the situation which we observe here.}

\section{Determination of the Eddington ratio}\label{app:medd}
The flat X-ray spectrum of \mcg\ ($\Gamma\approx1.7$ -- see Section~\ref{sec:xspec}) suggests a lower Eddington accretion ratio than previously reported for this AGN, $\dot{m}_\mathrm{E}\approx0.03$.
Here, we investigate the multiwavelength indicators of \me\ available for \mcg, assuming the RM mass $2.82\times10^7$~\msun, as detailed below.
The results are summarised in Table~\ref{tab:LLEdd}.

\subsection{Emission line estimates of $L_\mathrm{bol}$}\label{sec:nlr}
Narrow emission lines may be used to estimate the bolometric luminosities of AGN and the method has been applied to samples of type 2 AGN for which direct measurements of the accretion disc continuum cannot be made.
Because the narrow emission lines emanate from a large gaseous region beyond the AGN's dust torus they are less susceptible to extinction by dust in and around the central engine.
However, the large size of the emission region therefore means that the lines only give an indication of the Eddington ratio averaged over very long time frames.
From a fit to the mean optical spectrum of \mcg\ presented by \cite{Fausnaugh17} from the 2014 RM campaign, we determine $\log(L_{[\mathrm{O}\textsc{iii}]}/\mathrm{erg\,s}^{-1})=42.054\pm0.002$ and $\log(L_{\mathrm{H}\upbeta_\mathrm{n}}/\mathrm{erg\,s}^{-1})=41.20\pm0.02$.
(The luminosities have been corrected for Galactic dust extinction with $E(B-V)=0.18$: \citealt{SF11}.)
\cite{Lamastra09} provide a bolometric correction factor based on [O\,\textsc{iii}]~$\lambda5007$ for $\log(L_{[\mathrm{O}\textsc{iii}]}/\mathrm{erg\,s}^{-1})$ in the range 42--44.
From this factor and our measured $L_{[\mathrm{O}\textsc{iii}]}$, we derive an Eddington ratio of $\dot{m}_\mathrm{E}=0.14$.
(This is reduced to 5~per~cent if using the lower correction factor for $\log L_{[\mathrm{O}\textsc{iii}]}$ in the range 40--42.)
Using the lumninosity-dependent narrow H\,$\upbeta$ bolometric correction factor of \cite{Netzer19} and assuming a dust-free NLR, we again determine $\dot{m}_\mathrm{E}=0.15$.

We note, however, that both the \cite{Lamastra09} and \cite{Netzer19} bolometric conversions relate to line luminosities corrected for dust extinction within the NLR.
We estimate the NLR dust extinction for \mcg\ using the Balmer decrement.
\cite{Cohen83} gives the narrow-line flux ratio of H\,$\upalpha$ to H$\,\upbeta$ in \mcg\ as 5.40.
Assuming an intrinsic narrow-line Balmer decrement of 3.1 \citep{OF06}, we estimate that $E(B-V)\approx0.55$~mag is required to produce the observed ratio.
For a Milky Way type reddening curve (e.g.\ \citealt{CCM89}) the intrinsic line luminosities $L_{[\mathrm{O}\textsc{iii}]}$ and $L_\mathrm{H\,\upbeta}$ are a factor 5.8 and 6.2 higher than the observed ones, respectively (i.e.\ $\log(L_{[\mathrm{O}\textsc{iii}]}/\mathrm{erg\,s}^{-1})=42.817\pm0.002$ and $\log(L_{\mathrm{H}\upbeta_\mathrm{n}}/\mathrm{erg\,s}^{-1})=41.99\pm0.02$).
The estimated Eddington ratios are, of course, increased by the same factors. 

\subsection{X-ray estimates of $L_\mathrm{bol}$}
From our fit to the time-averaged \swift\ X-ray spectrum, combined with the 2021 \nustar\ spectra (Section~\ref{sec:xspec}), we determined the 2--10~keV X-ray continuum luminosity to be $\log(L_\mathrm{X}/\mathrm{erg\,s^{-1}})=43.68$.
Applying the \cite{Duras20} lumninosity-dependent bolometric correction factor we determine $\dot{m}_\mathrm{E}=0.23$.

The \swift\ BAT 157-month hard X-ray survey \citep{Lien25} reports a luminosity $\log(L_\mathrm{X}/\mathrm{erg\,s^{-1}})=44.13$ in the 14--195\,keV band.
This value is derived from the time-averaged spectrum generated from first 13 years of BAT observations (from 2004 December).
The estimation has the advantage that hard X-rays are not susceptible to moderate levels of photoelectric absorption.
\cite{Koss17} provide a simple bolometric correction factor $\kappa_\mathrm{bol}=8$ for BAT luminosities from which we determine $\dot{m}_\mathrm{E}=0.31$ although we note that this is an approximate correction, independent of the source luminosity.

\subsection{Infrared continuum estimates of $L_\mathrm{bol}$}
It is also possible to estimate the bolometric luminosity from infrared wavelengths which probe the large-scale dust torus surrounding the AGN central engine.
The dust readily absorbs and re-emits X-ray, UV and optical photons which make up the vast majority of an AGN's radiative output.
The dust emission is therefore an effective bolometer for the AGN central engine.
Because the torus is a very large scale structure ($\gtrsim$~pc or $\gtrsim$~3\,light years), the derived bolometric luminosity is an average over a long time scale (up to $\sim$decades).

We queried the \textit{Wide-field Infrared Survey Explorer} (\textit{WISE}; \citealt{Wright10}) source catalog and the 2MASS All-sky Point Source Catalog (\citealt{Cutri03}; \citealt{Skrutskie06}), both hosted on the Infrared Science Archive\footnote{\url{https://irsa.ipac.caltech.edu/}}, to obtain IR magnitudes of \mcg.  
The \textit{WISE} photometry relates to observations taken during the main survey (2009 December -- 2010 August); the 2MASS survey was conducted between 1997 June and 2001 February.
The IR photometric magnitudes are listed in Table~\ref{tab:wise}.

\begin{table}
    \centering
    \caption{Archival IR photometry}
    \begin{tabular}{cccc}
    \hline
    Filter               & $\lambda$        & Magnitude      & $\lambda F_\lambda$ \\
                         & [\si\um]         &                & [$10^{-11}$~erg\,s$^{-1}$\,cm$^{-2}$] \\
    \hline
    2MASS J              &  1.24            & $12.33\pm0.05$ & $5.7\pm0.3$  \\
    2MASS H              &  1.66            & $11.29\pm0.05$ & $6.6\pm0.3$  \\
    2MASS K              &  2.16            & $10.24\pm0.03$ & $8.2\pm0.2$ \\
    \textit{WISE} W1     &  3.36            &  $8.36\pm0.02$ & $15.1\pm0.2$ \\
    \textit{WISE} W2     &  4.60            &  $7.34\pm0.02$ & $13.7\pm0.2$ \\
    \textit{WISE} W3     & 11.6             &  $4.39\pm0.01$ & $13.0\pm0.2$ \\
    \textit{WISE} W4     & 22.1             &  $1.76\pm0.01$ & $23.2\pm0.8$ \\
    \hline
    \end{tabular}
    \label{tab:wise}
\end{table}

We derive infrared luminosities from the \textit{WISE} W3 (12\,\si\um) and W4 (22\,\si\um) magnitudes, and the K-band (2\,\si\um) magnitude from 2MASS.
We use the luminosity-dependent infrared bolometric corrections of \cite{Runnoe12} (see Table~\ref{tab:LLEdd}) and determine Eddington ratios which decrease with wavelength from 0.96 to 0.21.
It should be noted that the infrared magnitudes have not been corrected for any host galaxy contamination.
Although it has been noted that for quasars the infrared emission from the host galaxy typically becomes dominant only at wavelengths longer than 24\,\si\um\ (\citealt{Runnoe12} and references therein), the same may not be true for Seyferts such as \mcg\ which have much lower AGN luminosities.
The apparent increase of the Eddington ratio with wavelength may then be attributed to increasing host galaxy contribution to the infrared flux.
We may then consider the shorter-wavelength (K and J) IR estimates of $\dot{m}_\mathrm{E}\sim0.20$ to be the more reliable. 

\begin{table*}
    \centering
    \caption{Various estimates of the Eddington ratio $\dot{m}_\mathrm{E}$}
    \begin{tabular}{clcccc}
    \hline
    Estimator                             & Bolometric correction   & $\kappa_\mathrm{bol}$ & $\log(L_\mathrm{obs})$ & $\log(L_\mathrm{bol})$ & $\dot{m}_\mathrm{E}=L_\mathrm{bol}/L_\mathrm{Edd}$ \\
    \hline
    22.19~\si\um\ (W4)                    & \cite{Runnoe12} eqn.\ 8 & 16 & 44.33 & 45.53 & 0.96 \\
    12.08~\si\um\ (W3)                    & \cite{Runnoe12} eqn.\ 6 & 11 & 44.14 & 45.20 & 0.45 \\
     2.16~\si\um\ (K)                     & \cite{Runnoe12} eqn.\ 3 & 10 & 43.85 & 44.86 & 0.21 \\
     1.24~\si\um\ (J)                     & \cite{Runnoe12} eqn.\ 2 & 22 & 43.63 & 44.98 & 0.27 \\
    H\,$\upbeta_\mathrm{n}$~$\lambda4861$ & \cite{Netzer19} table 1 & 3288 (4561) & 41.20 (41.99)  & 44.72 (45.64)  & 0.15 (1.23) \\
    {[O\,\textsc{iii}]}~$\lambda5007$     & \cite{Lamastra09}       & 454  & 42.054 (42.817) & 44.71 (45.47) & 0.14 (0.83) \\
    2--10 keV                             & \cite{Duras20} table 1  & 17   & 43.68  & 44.91   & 0.23  \\
    14--195 keV                           & \cite{Koss17}           &    8 & 44.13  & 45.03   & 0.31 \\
    \hline
    \end{tabular}
    \parbox{13.5cm}{Luminosities are in units of erg\,s$^{-1}$.  
    For an assumed mass of $2.82\times10^7~\mathrm{M}_\odot$ the Eddington luminosity is $\log(L_\mathrm{Edd}/\mathrm{erg\,s^{-1}})=45.55$.
    For H\,$\upbeta_\mathrm{n}$ and [O\,\textsc{iii}] numbers in parentheses include a correction for dust extinction in the NLR, as described in Section~\ref{sec:nlr}.
        }
    \label{tab:LLEdd}
\end{table*}

\bsp	
\label{lastpage}
\end{document}